\documentclass[preprint,3p,12pt]{elsarticle}
\usepackage{color}
\usepackage{amsfonts}
\usepackage{amsmath}
\usepackage{graphicx}
\usepackage{epstopdf}
\usepackage{fouriernc}
\usepackage{amssymb}

\journal{ }
\begin{document}
\begin{frontmatter}
\title{Reactive strip method for mixing and reaction in two dimensions }
\author[1]{Aditya Bandopadhyay}
\ead{aditya.bandopadhyay@univ-rennes1.fr}
\author[1]{Tanguy Le Borgne}
\author[1]{Yves M\'eheust}
\address[1]{Universit{\'e} de Rennes 1, CNRS, G{\'e}osciences Rennes UMR 6118, 35042 Rennes, France}

\begin{abstract}

A numerical method to efficiently solve for mixing and reaction of scalars in a two-dimensional flow field at large P{\'e}clet numbers but otherwise arbitrary Damk\"ohler numbers is reported. Flow disorder often leads to the formation of lamellar structures for the reactants, thus altering the observed reaction rates.
We consider a strip of one reactant in a pool of another reactant, both of which are advected with a known velocity field. We first establish that the conditions under which the system may be described by a locally one-dimensional reaction-diffusion problem is when the strip thickness is smaller than the local radius of curvature and also when the strip thickness is smaller than the distance between adjacent strips. 
In such a scenario, typical of many mixing systems, the system of advection-diffusion-reaction equation in two-dimensions is thus reduced to a local one-dimensional reaction-diffusion equation in the Lagrangian frame attached to the advected strip in strip in such a manner that the effect of advection is systematically decoupled  from the diffusion and reaction processes. 
We first demonstrate the method for the transport of a conservative scalar (the limiting case of zero Damk\"ohler number) under a linear shear flow,  point vortex and a chaotic sine flow. We then proceed to consider the situation of a simple bimolecular reaction between two reactants yielding a single product. 
In essence, the reduction of  dimensionality of the problem, which renders the 2D problem 1D, allows one to efficiently model reactive transport under high P{\'e}clet numbers which are otherwise  prohibitively difficult to resolve from classical finite difference or finite element based methods. 

\end{abstract}

\begin{keyword}
Mixing and reaction, high P\'eclet number
\end{keyword}

\end{frontmatter}

\section{Introduction} \label{sec:intro}

The flow of reactive species plays a critical role in several aspects of industrial processes, geophysical and atmospheric processes as well as biological processes \cite{tel2005chemical,kitanidis2012delivery}. From an industrial point of view, mixing and reaction in fluids, be it in a batch process or a continuous flow process, is central to several processing operations. For several important industrial and chemical processes such as plastic molding, paint processing, food and pharmaceutical processing, paper processing etc. which are inherently highly dissipative in nature owing to the relatively stronger viscous forces as compared to the inertial forces, it is often the best strategy to mechanically stir, stretch, and wrap the multiple species so as to amplify the concentration gradients \cite{danckwerts1958,danckwerts1953,mohr1957,ottino1989mixing,shankar2009,finn2011,villermaux2000}.

Processes such as CO$_2$ sequestration and ozone layer dynamics examples of such reactive flows. For mixing and reaction in atmospheric flows (for example, the ozone layer depletion by chlorine monoxide), the process is often modeled as a 2D stirring flow \cite{wonhas2002diffusivity}. There exist several instances where  fluid stretching and reaction has a significant impact on the overall reaction rate and product formation rate \cite{boano2007bedform,boano2006sinuosity,toth1963theoretical,steefel2009fluid,maher2014hydrologic}. The analysis of such processes is also central towards understanding the formation of biofilms, biostreamers and interaction amongst multibacterial colonies in flows through reactors or porous media \cite{stoodley1999formation,characklis1981bioengineering,stocker2012marine,ebrahimi2016microbial}. 
For example, it was shown by Birch et al. \cite{birch2008thin} that plankton dynamics (birth and growth) in oceanic currents leads to the formation of a thin layer of plankton (whose birth and death is dictated by nutrient uptake) acted upon by a linear shear flow. Martin \cite{martin2003phytoplankton} and Abraham \cite{abraham1998generation} have also demonstrated that the blooms of phytoplankton are indeed well described as a transport of reactive species in pseudo-two dimensions (referred to as patchiness of the bloom) -- as observed through  Advanced Very High Resolution Radiometer (AVHRR) imaging. Laminar mixing by means of geometric features in the flow channels is nowadays employed in the paradigm of micromixing wherein the low Reynolds numbers usually leaves very little scope for achieving good mixing through only molecular diffusion \cite{stone2004,kim2004,yang2005,villermaux2008,cortes2009,amini2013}. On the other side of the spectrum away from microfluidics, data from field tracer tests are central towards the analysis of fractures and distribution of porosity of a porous media connecting two reservoirs which may span several tens of meters \cite{burr1994nonreactive}. 

Mixing has received attention from several researchers owing to the ubiquity of the problem \cite{ottino1990mixing}. Franjione and Ottino \cite{franjione1989symmetries,franjione1991stretching} have considered the problem of the stretching of material lines in 3D duct flows and in confined 2D flows. Through a series of papers, Beigie et al. \cite{beigie1991chaotic,beigie1991global,beigie1993statistical} have addressed through rigorous mathematics the influence of stretching on the dispersion of a scalar in confined 2D forced flows. One of the key assumptions for the analytical tractability of these problems is that diffusion is essentially 1D (confined to the direction of the largest gradient) for the stretched segments. Clifford et al. \cite{clifford1999reaction} have studied the effect of the formation of lamellar structures during fluid mixing towards assessing the increase in the net reaction rate. Sokolov and Blumen have studied analytically the asymptotic behaviour of such a periodic lamellar structure for the evolution of concentration of reactants \cite{sokolov1991diffusion}. The limiting case of fast binary reactions lends itself to analytical tractability -- and has been a topic of study in recent times as well \cite{de2005procedure}. The influence of fluid material line stretching has been analyzed for the case of fast bimolecular reactions \cite{borgne2014impact}. Besides this, Zolt\`an and Hern{\'a}ndez-Garc{\'\i}a have discussed the influence of a 2D straining flow on the nature of bimolecular reactions observed at the front demarcating two reactants which are reacting infinitely fast \cite{emilio}.

In the case of reactive gaseous species, one may exploit turbulent flow conditions which lead to a very rapid rate of mixing. Based on the ideas laid by Batchelor \cite{batchelor1952effect,batchelor1952diffusion} on the stretching of material lines and scalar diffusion, Marble and Broadwell \cite{marble1977coherent,marble1988mixing} attempted to model the process of combustion and chemical reactions using the coherent flame model. Ranz and co-workers also worked on modeling how the stretching of material lines is responsible for enhancing mixing through diffusion \cite{ranz1979,ottino1979} and how it affects the rate of consumption of one species \cite{ou1983}. These were the earliest instances for which researchers actively considered the effect of the imposed fluid flow on overall stretching and compression process rather than just obtaining the  residence time of a given material in the flow. In the context of turbulence flow, the various aspects of mixing in a Lagrangian framework have been reviewed extensively by Yeung \cite{yeung2002lagrangian} and Vassilicos \cite{vassilicos2002mixing,wonhas2001mixing}. The influence of external perturbations on such stirred chaotic flows -- termed as excitable flows (employing the chaotic sine flow, briefly discussed later) has also received attention in recent times \cite{neufeld2001excitable,neufeld2002excitable,hernandez2003filament}.

It is clear that modeling such flows is therefore of paramount importance to efficiently understand the intricate structure and dynamics of the aforementioned processes.  Eulerian methods are relatively straightforward but require great computational power to fully resolve the solution grid and time steps for flows with high P{\'e}clet number. In light of this  there have been active efforts to depict the problem of reactive transport, given the complete knowledge of the flow field, in a simplified framework. In the case of high P{\'e}clet numbers,  weak diffusion effects in comparison to advection effects lead to the species being primarily confined in the vicinity of the material line that has advected that particular species. In fact in the limit of nondiffusing species,  the species  follow the fluid material lines \textit{exactly}. Therefore the approach that we develop here is to transform the equations in 2D by making appropriate coordinate transforms to the so-called \textit{warped coordinates} \cite{ranz1979, emilio}. By virtue of these coordinate transforms, we are able to account for the advection by means of the kinematic relations. Then on these advected material lines, we can then account for diffusion and reaction separately from advection.  In the case of conservative scalars, this methodology has been shown to be extremely efficient in predicting all the quantities of interest in not just in a qualitative manner but also in a quantitative manner \cite{meunierDSM}. It is in this context that we aim to generalize the case of a conservative scalar to a system of reactive species.

\section{Mathematical formulation} \label{sec:math}
\subsection{2D formulation} \label{sec:2D}
Let us consider the advection-diffusion-reaction system comprising  two reactants $A$ and $B$ to yield a product species $C$ whose concentrations are denoted by $a$, $b$ and $c$ respectively. For simplicity, we may assume here that the diffusion coefficients of the species are equal and denoted by $D$. With this consideration, we may write the governing equations for the transport of $A$, $B$, and $C$ as 
\begin{equation}
\begin{gathered}
\frac{{\partial \tilde a}}
{{\partial \tilde t}} + {\mathbf{\tilde u}}\cdot\nabla \tilde a = D{\nabla ^2}\tilde a - \tilde{R}, \hfill \\
  \frac{{\partial \tilde b}}
{{\partial \tilde t}} + {\mathbf{\tilde u}}\cdot\nabla \tilde b = D{\nabla ^2}\tilde b - \tilde{R}, \hfill \\
  \frac{{\partial \tilde c}}
{{\partial \tilde t}} + {\mathbf{\tilde u}}\cdot\nabla \tilde c = D{\nabla ^2}\tilde c + \tilde{R}, \hfill 
\end{gathered}
\label{eq:gov_eq}
\end{equation}
where the variables with tilde represent   dimensional quantities and $\tilde{R}$ represents the rate of reaction whose form is not yet prescribed. Although the methodology presented herein is valid for any general expression for the reaction kinetics, let us, for simplicity, consider the case of a second order bimolecular reaction for which we may represent the reaction rate as $R = k\tilde{a}\tilde{b}$. Let us proceed to nondimensionalize the time as per the advection timescale $t_\text{ref} = L/U$ where $L$ represents any pertinent characteristic lengthscale of the system and $U$ represents a reference velocity scale of the system. Let $n_0$ be a characteristic scale for the species concentration that is appropriate in the context of the system under consideration. With these considerations, we may write equation \eqref{eq:gov_eq} in a nondimensional form as 
\begin{equation}
\begin{gathered}
  \frac{{\partial a}}
{{\partial t}} + {\mathbf{u}}\cdot\nabla a = \frac{1}
{{Pe}}{\nabla ^2}a - {{Da_\text{II}}}
\,ab, \hfill \\
  \frac{{\partial b}}
{{\partial t}} + {\mathbf{u}}\cdot\nabla b = \frac{1}
{{Pe}}{\nabla ^2}b - {{Da_\text{II}}}
\,ab, \hfill \\
  \frac{{\partial c}}
{{\partial t}} + {\mathbf{u}}\cdot\nabla c = \frac{1}
{{Pe}}{\nabla ^2}c + {{Da_\text{II}}}
\,ab, \hfill \\ 
\end{gathered}
\label{eq:nd_gov_eq}
\end{equation}
where the P{\'e}clet number, $Pe=UL/D$, appearing in equation \eqref{eq:nd_gov_eq} represents the ratio of the diffusion timescale to the advection timescale and is given by $Pe = UL/D$; the Damk{\"o}hler number of the second kind, $Da_\text{II} = kn_0L/U$ which quantifies the ratio of the typical advection time scale to the reaction time scale is basically the ratio between the Damk{\"o}hler number of the first kind, $Da$ (which quantifies the ratio of the time scale for diffusion to the time scale of reaction and is given by $Da = kn_0L^2/D$) and the P\'eclet number $Pe$. 

The advection-diffusion-reaction equation \eqref{eq:nd_gov_eq} is common to many scalar transport equations which may have a source or sink, signifying the reaction and/or consumption of the interacting scalars. 
As was earlier pointed out, owing to typical stirring flows occurring at high P{\'e}clet numbers, the main drawback of simulating the set of equations \eqref{eq:nd_gov_eq} remains in the immensely fine spatial resolution needed to capture filamentous structures and the small time steps required to prevent numerical instabilities \cite{yeung2005high}. 

A few comments are in order about the set of transport equations for the three species as depicted in equation \eqref{eq:gov_eq}. The choice of the dimensionless velocity, length and time are based on the advection scale. This allows us to track the material lines in a velocity field which is $\sim O(1)$. This particular choice of scales leads to the diffusion equation being multiplied by a prefactor $1/Pe$. For large $Pe$ it is clear that there will be a diffusion boundary layer near the filaments. The width of this diffusion boundary layer, considering the case where $Da_\text{II}$ is small (since the axial direction contributes to diffusion to a very small extent in comparison to the transverse/perpendicular direction), is expected to scale as $Pe^{-1/2}$ \cite{Leal,kundu2008fluid}. Thus, as the P\'eclet number increases, the boundary layer becomes even thinner which allows the premise of the present method, that the species concentrations remain confined to regions near the strip, to hold true. The second point is the typical values  of $Da_\text{II} = Da/Pe$ for finite reaction rates. For the cases where $Da \ll Pe$, which corresponds to the kind of flows that we hope to address through this work -- high P\'eclet number flow, the parameter $Da_\text{II}$ is observed naturally  to be very small.

\subsection{Lagrangian frame formulation} \label{sec:1D}

\begin{figure}[!ht]
\begin{centering}
\includegraphics[scale=0.6]{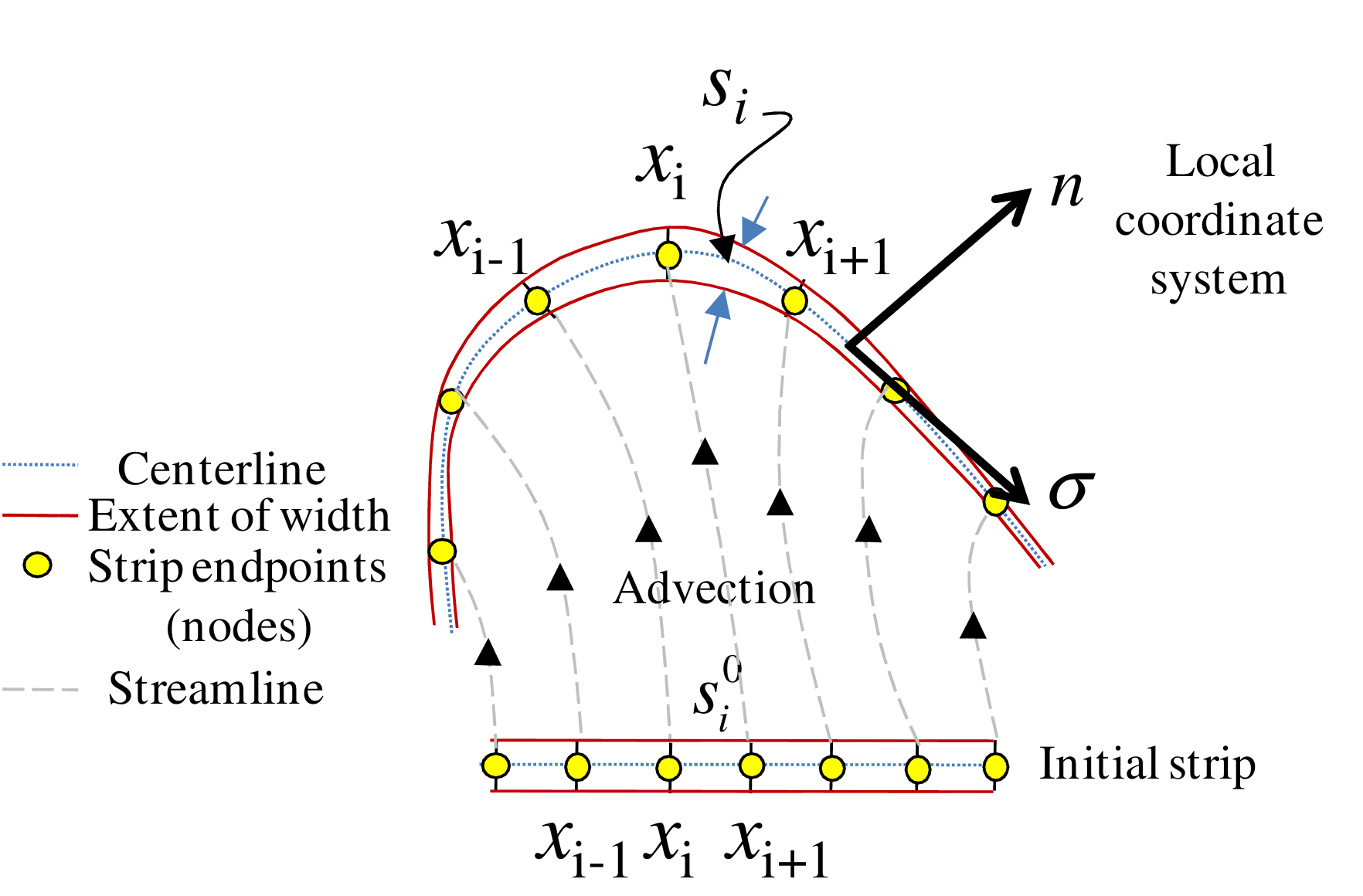}
\caption{Schematic of the representation of a material line as a set of discretized strips. The initial strip of one reactant is placed in the bulk of another reactant. In the schematic, the solid lines highlight the width of the initial strip. The dashed lines indicate streamlines for the flow field. While being advected, the conservation of area implies that the area of the material volume is conserved. The content of this material volume diffuses and reacts while simultaneous being advected in the flow. Consequently, there is a thinning (or thickening) of the width of the reactant. Alternatively, the process can also be depicted in terms of the width of the product formed. The initial thickness of the strip is $s_i^{0}$ and as the strips are stretched or compressed the strip thickness transforms to $s_i$ depending on which the gradient for diffusion decreases or increases. If the stretched distance is larger than the thickness of the strip, the problem may be treated as essentially locally 1D, with the diffusion and reaction primarily occurring in the $n$ direction.}
\label{Fig:schematic}
\end{centering}
\end{figure}

To avoid  problems due to the description of filamentous structures and fine structures in stirring flows, we aim at simplifying the governing equations in 2D (equations \eqref{eq:nd_gov_eq}) into a 1D scenario. Towards this, we follow the framework similar to that proposed by Meunier and Villermaux, who considered the problem for a conservative scalar \cite{meunierDSM,bandopadhyay2016enhanced}. 

The fundamental concept of reducing of the dimensionality of the problem from 2D to 1D was laid by Ranz in his seminal paper \cite{ranz1979}. It was shown that the material line is stretched in the primary directions of the flow, deforming it in one direction and compressing it in the other direction. In such a case, the gradients in the directions parallel to the structure so formed can be neglected and the scalar transport is simply advection with the flow field (and the material line) and this is accompanied by diffusion and reaction in the perpendicular direction. Therefore, only the direction normal to the stretching material line/interface is enough to yield the necessary information about the entire 2D process. 

We may discretize a long continuous material line into several strips for convenience for further numerical computations and evaluations. For each of these discretized strips, we therefore have a definite velocity field dependent stretching. 
We assume that the flowfield ${\bf{u}}({\bf{x}},t)$ is known, either analytically, numerically or from experimental measurements. 
Implicit in this assumption is that the velocity field is decoupled from the concentration field. The concentration field is, however, advected by the flowfield. We consider a domain of lengths $\mathcal{\tilde{L}}\times\mathcal{\tilde{H}}$ which, in a nondimensional sense, may be written as $\mathcal{L}\times\mathcal{H}$ where the pertinent characteristic lengthscale, $L$, is appropriately chosen as per the discussion after equation \eqref{eq:ranz}. In this domain we consider a strip of a given length $\tilde{l}$ (nondimensional length $l$) and width $\tilde{s}$, (nondimensional width $s$). 

Owing to the action of a 2D flow, the original strip deforms and consequently is elongated in one direction while it gets compressed in the other direction due to the constraint of incompressibility. In other words, for the material line, we may represent, for a strip, the condition of incompressibility in the form of an area conservation equation (please refer to figure \ref{Fig:schematic})
\begin{equation}
s_i = s_0 \frac{ l_i^0}{ l_0}
\label{eq:kinematic}
\end{equation}
where $l_i^0$ and $s_i$ represent the initial length and thickness of the segment $i$. 
With the assumption that the axial gradient (which refer to the direction oriented along the strip, $\sigma$) can be neglected, the dimensionless approximate equations for the transport of the three species may be written as (refer to  \ref{sec:Appendix_A}) \cite{Meunier2003,meunierDSM}

\begin{equation}
\begin{gathered}
  \frac{{\partial a}}
{{\partial t}} + \frac{1}
{{{s_i}}}\frac{{d{s_i}}}
{{dt}}n\frac{{\partial a}}
{{\partial n}} = \frac{1}
{{Pe}}\frac{{{\partial ^2}a}}
{{\partial {n^2}}} - Da_\text{II}\,ab, \hfill \\
  \frac{{\partial b}}
{{\partial t}} + \frac{1}
{{{s_i}}}\frac{{d{s_i}}}
{{dt}}n\frac{{\partial b}}
{{\partial n}} = \frac{1}
{{Pe}}\frac{{{\partial ^2}b}}
{{\partial {n^2}}} - Da_\text{II}\,ab, \hfill \\
  \frac{{\partial c}}
{{\partial t}} + \frac{1}
{{{s_i}}}\frac{{d{s_i}}}
{{dt}}n\frac{{\partial c}}
{{\partial n}} = \frac{1}
{{Pe}}\frac{{{\partial ^2}c}}
{{\partial {n^2}}} + Da_\text{II}\,ab, \hfill \\ 
\end{gathered} 
\label{eq:citrakis}
\end{equation}
where the only derivatives present in the governing equation are in the direction normal to the direction of the strip, $n$. 
Therefore, equation \eqref{eq:citrakis} represents a dramatic simplification of the full governing equations for species transport from a 2D representation to a 1D representation for each strip $i$. 
We now aim at reducing the formalism further into a diffusion-reaction equation by changing our viewpoint from the Eulerian frame to the Lagrangian frame attached to the strip. 

To begin, we consider the strip in the Lagrangian frame. In this frame we make a change of coordinate. We may refer to these as the warped coordinates (we are referring to the local coordinate system in figure \ref{Fig:schematic}).
The warped time, $\theta$, is given as the integral diffusion time scale and is defined as 
\begin{equation}
\frac{d\theta}{dt} = \frac{1}{Pe\, s^2},
\label{eq:warped}
\end{equation}
On similar lines, the rescaled spatial coordinate in the normal direction of the strip, $z$, is given as 
\begin{equation}
z=\frac{n}{s(t)}.
\label{eq:rescaledz}
\end{equation}
Using these warped coordinates, we may simplify equation \eqref{eq:citrakis} further to obtain \cite{meunierDSM,bandopadhyay2016enhanced,emilio}

\begin{equation}
\begin{gathered}
  \frac{{\partial a}}
{{\partial \theta }} = \frac{{{\partial ^2}a}}
{{\partial {n^2}}} - Da_\text{II}Pe\,ab\,{s^2}, \hfill \\
  \frac{{\partial b}}
{{\partial \theta}} = \frac{{{\partial ^2}b}}
{{\partial {n^2}}} - Da_\text{II}Pe\,ab\,{s^2}, \hfill \\
  \frac{{\partial c}}
{{\partial \theta}} = \frac{{{\partial ^2}c}}
{{\partial {n^2}}} + Da_\text{II}Pe\,ab\,{s^2}, \hfill \\ 
\end{gathered} 
\label{eq:ranz}
\end{equation}

Equation \eqref{eq:ranz} is  a convenient framework for representing the transport of the reactants and product in the Lagrangian framework. It must be noted that the effect of advection is solely manifested in the terms of an altered reaction term $Da\; ab\;s^2$. 
It may be noted that the transport in the presence of reactions is fundamentally different from the transport of a conserved scalar in the sense that the reaction rate depends on the history of the stretching -- as observed from the contribution of the $s^2$ term in the reaction term. 
For a conserved scalar, the time evolution in the $\{\theta,z\}$ space merely depends solely on the final warped time $\theta$ of a particular strip (please refer to \ref{sec:Appendix_A}) whereas in the case of chemical reactions, the evolution history, which may be often complicated, involves the knowledge of the amount of stretching that an interface has encountered during the evolution of the system and this stretching history continuously affects the rate of reactions seen in the system (please refer to the right hand side of the governing equations for the concentration fields \eqref{eq:ranz}). It is precisely the fact that we must keep a track of the deformation history of each of the strip that renders the problem significantly more complex than the  case of a conservative scalar. 

Clifford et al. \cite{clifford1998lamellar} derived similar equations for a local 1D reactive transport for the case of a chaotic flow. For a chaotic flow the rate of stretching or mixing is quantified entirely in terms of an exponential stretching rate. They argue that since the strips are stretched exponentially fast, one may drop the axial diffusion terms and only retain the advection and diffusion in the direction perpendicular to the strip. After this, the coordinate in the perpendicular direction may be rescaled through $\exp(-\alpha t)$ where $\alpha$ represents the average stretching rate. We must point out that these results are representative of a special case of stretching and are not applicable for a general case. In general one may also have a linear rate of linear or algebraic stretching \cite{le2015lamellar,duplat2010nonsequential}. As a matter of fact, the nature of stretching may be different for different regions in the flow field. 
We also note that in order to arrive at the same equations obtained for this special case of exponential stretching, we may simply substitute $s = \exp(-\alpha t)$. It may be observed that the form of warped time for an exponential stretching is also an exponential $\theta = \frac{1}{2\alpha Pe} \exp(2 \alpha t) $.

\section{Numerical method} \label{sec:nummeth}

\begin{figure}[!ht]
\begin{centering}
\includegraphics[scale=0.5]{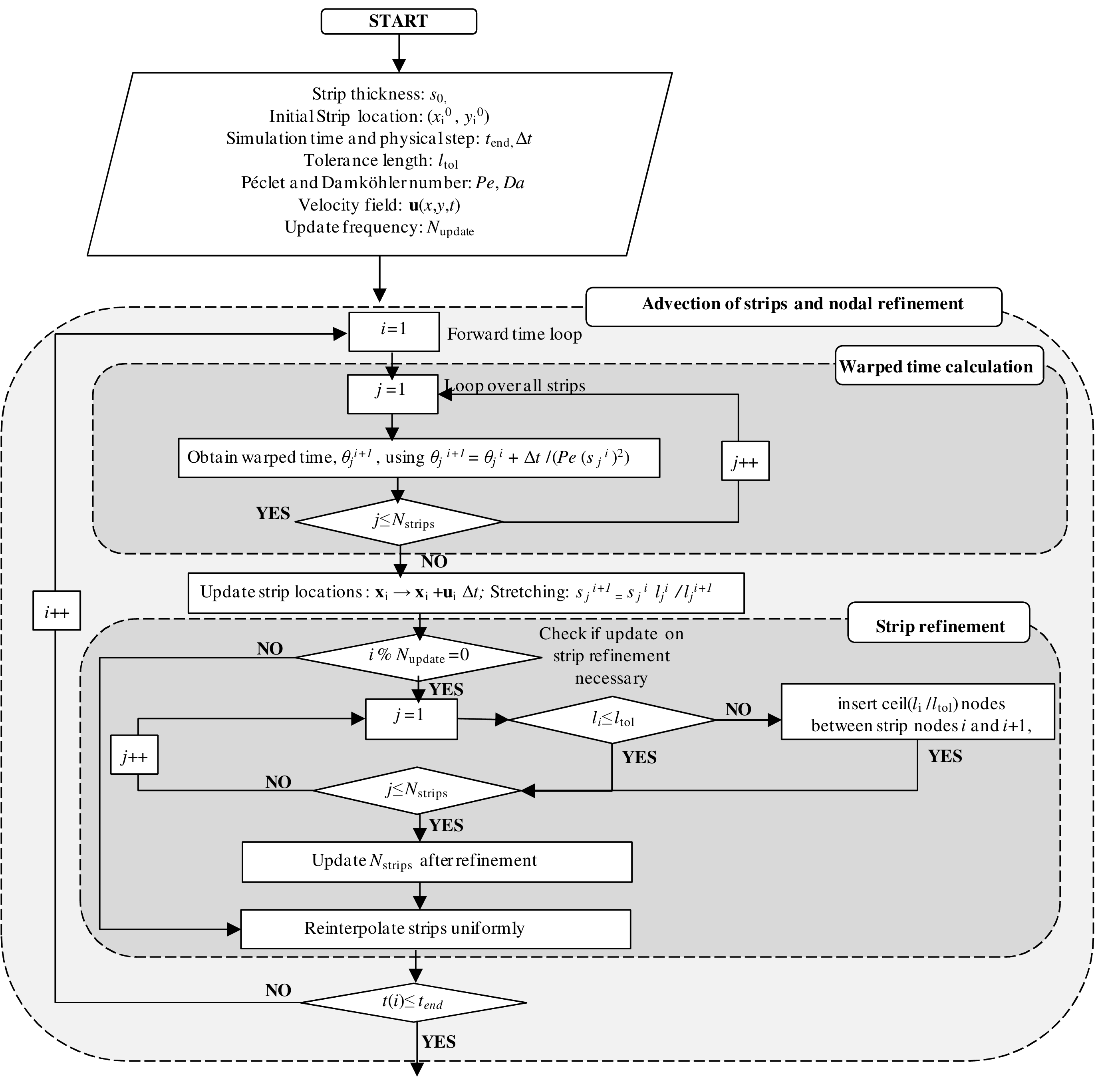}
\caption{Flowchart for the first step of the algorithm of the Reactive Strip Method. The first step is advection of the curves in a specified flow field with a dynamic node insertion subjected to tolerances on the distance between two nodes. This algorithm then continues to the backiteration process described through the flowchart in figure \ref{Fig:flowchart2}.}
\label{Fig:flowchart1}
\end{centering}
\end{figure}

We split the overall method into a number of substeps, which we discuss in this section.

\subsection{Evaluating warped time}
The most fundamental difference between the formulations in the Eulerian frame and in the Lagrangian frame is the description of the time over which the process takes place. We have already shown in equation \eqref{eq:ranz} that the temporal variable is now converted into the warped time (defined through equation \eqref{eq:warped}). In general, in a nonuniform flow, the warped time is different for different strips. For extensively stretched strips the warped time is much larger (on account of the reduction of the width of the strip associated with the longitudinal extension). The larger warped time reflects the fact that  there is a enhancement of diffusion on account of larger extension of the strips. 
 
At each time step, the warped time for a given strip may be evaluated by integrating equation \eqref{eq:warped} in time using a simple forward difference scheme 
\begin{equation}
\theta_i \rightarrow \theta_i + \Delta t/(Pe\times s_i^2)
\label{eq:warpedtime}
\end{equation}
subjected to the initial condition that at $t = 0$ we must have $\theta_i = 0 \;\forall \;i$.
We have represented the flowchart explaining the kinematic aspects of the Reactive Strip Method though the flowchart depicted in figure \ref{Fig:flowchart1}. The calculation of the warped time represents the first substep of the broader step of the advection and refinement of strips. A simple loop (implemented vectorially in MATLAB) is sufficient to obtain the warped time for all the strips at the next time level. 

\subsection{Advection of strips and nodal refinement}
\label{sec:fwd}
This is the purely kinematic step which encompasses the information about the flow by advecting strips along with the flow. 
We begin by focusing only one strip $s_i$ which is described by the points $[x_i, x_{i+1}]$ and $(y_i,y_{i+1})$. The initial strip thickness represented as $s_i^0$. In the known flow field, we march the particles in time with a timestep given by $\Delta t$. With this, we may update the location of the ends of the strip through the passive advection of the points given by 
\begin{equation}
x_{i}\rightarrow x_{i}+u(x_{i},y_{i},t)\Delta t; \; y_{i}\rightarrow y_{i}+v(x_{i},y_{i},t)\Delta t 
\end{equation}
This step is clearly elucidated in figure \ref{Fig:schematic} which shows how the initial material strips are purely advected in the flow field. This advection step represents the precursor to the strip refinement process, which we shall elaborate below. 

\begin{figure}[!th]
\begin{centering}
\includegraphics[scale=0.8]{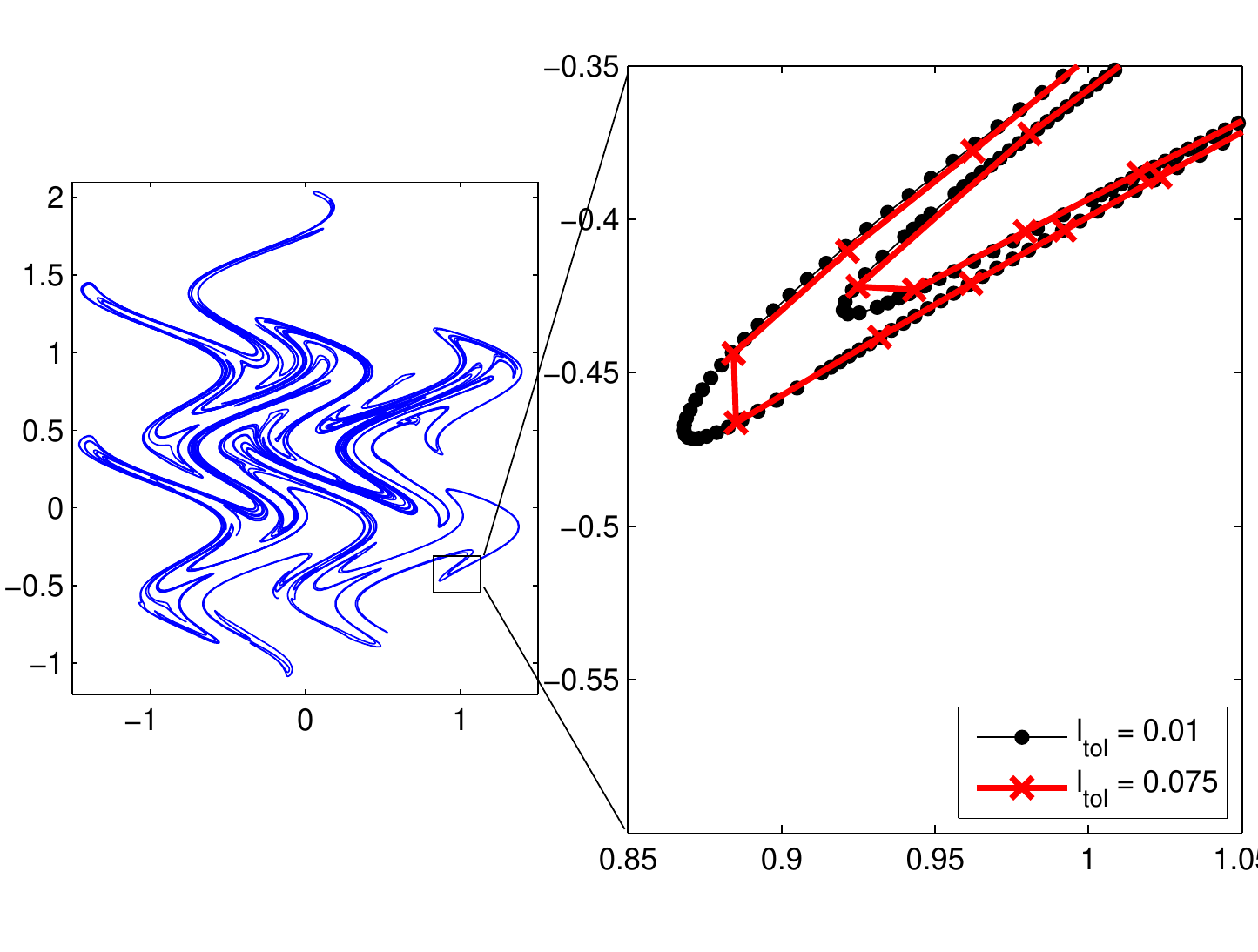}
\caption{Distribution of a strip under subjected to a chaotic sine flow at $t = 7$ that was initially distributed along the $y$ axis from -0.5 to 0.5 . The magnified portion of the strip is depicted for two different  length tolerances $l_{tol} = $(a) 0.01 and (b) 0.075. One can clearly observe the higher nodal density in the region with the bend in comparison to the straighter segment. The reconstruction is, quite obviously, better for the case with a smaller $l_{tol}$.}
\label{Fig:refinement}
\end{centering}
\end{figure}

Quite obviously, for a faithful representation of a \textit{continuum} material line one would ideally require a very large number of initial points. In this way, sharp flow reversals and gradients would be accurately captured and the shape formed by connecting the different nodes (whose coordinates in 2D is represented by means of $x_i$ and $y_i$) would be a fair representation of a purely advective species (non diffusing scalar). However, an \textit{a priori} fixed number of nodes does not seem appropriate as there may be a large fraction of the flow field where the gradients are not severe. In such regions it is sufficient to have a certain lower number of nodes so as to represent the stretching material line with reasonable accuracy without a significant memory overhead -- a significant issue as the length of the strip increases rapidly, in a chaotic flow for example.
In conclusion, it may be noted here that in the case of highly stretching flows, it may so happen that the distance between the two points increases to such an extent that the straight line connecting the two end points defining the strip is not a good indicator of the actual dynamics of the strip because of the lack of resolution in capturing the gradients between the endpoints of the strips. 
This aspect has been a topic of study in the area of nonlinear dynamics of chaotic flows such as the sine flow. 
In such kinds of flow, it is perhaps most efficient to start off a given advection run with a modest number of nodes and have a dynamic refinement of the strips, in which  points are inserted based on the condition of internodal separation or curvature \cite{alvarez1998self,meunierDSM}.

\begin{figure}[!th]
\begin{centering}
\includegraphics[scale=1]{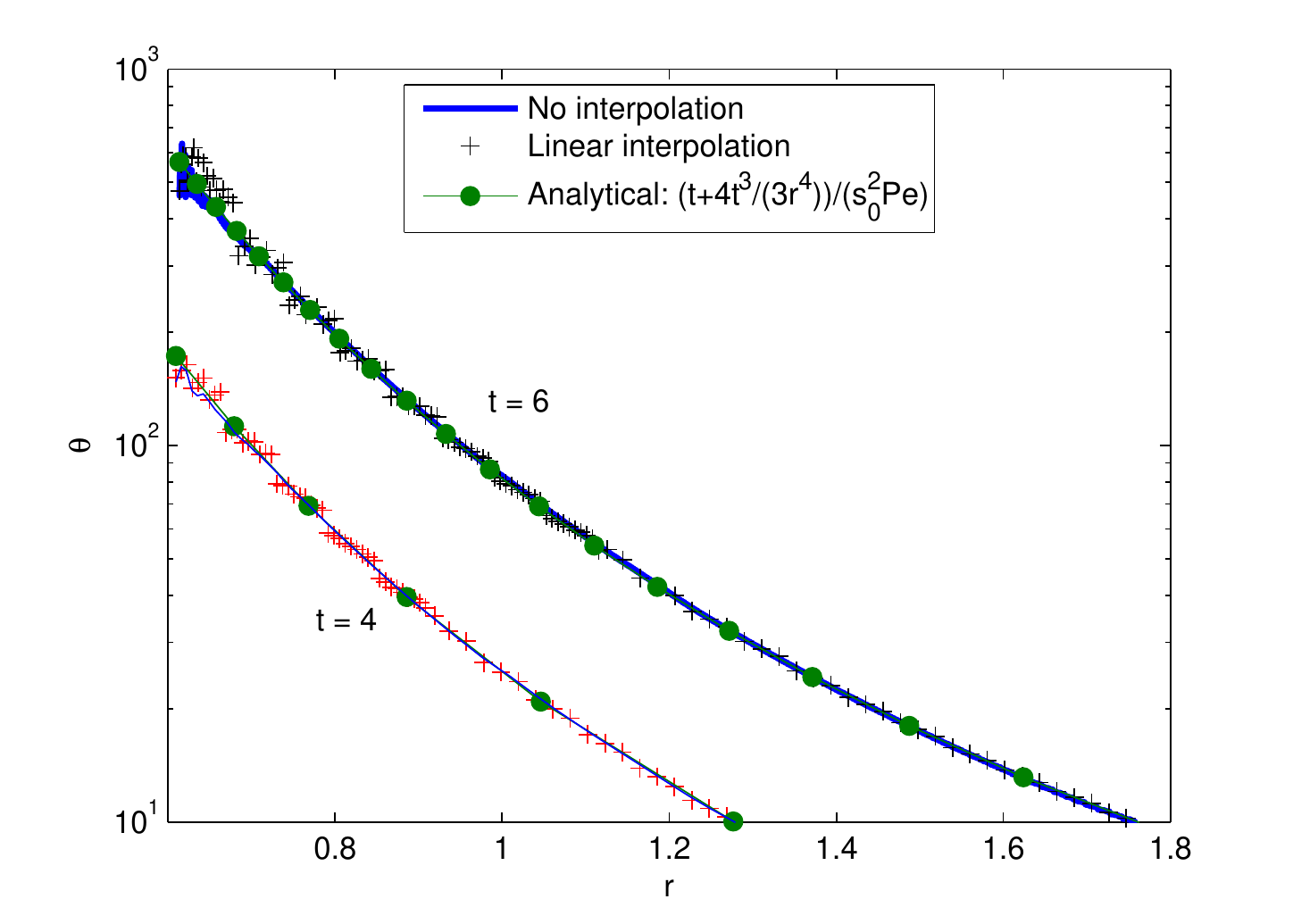}
\caption{The influence of the strip reinterpolation on the evaluation of the warped time for a point vortex flow. It is seen that the influence of the reinterpolation of the strip yields much better predictions of the warped time than without interpolation. Without reinterpolation, there is significant fluctuation of the warped time about the mean (the mean is, interestingly close to the analytical result). For the figure, we have chosen a tolerance of 0.02 and new points are inserted into the strip and reinterpolated every 20 steps.}
\label{Fig:interpolationerror}
\end{centering}
\end{figure}

\begin{figure}
\begin{centering}\
\includegraphics[scale=0.5]{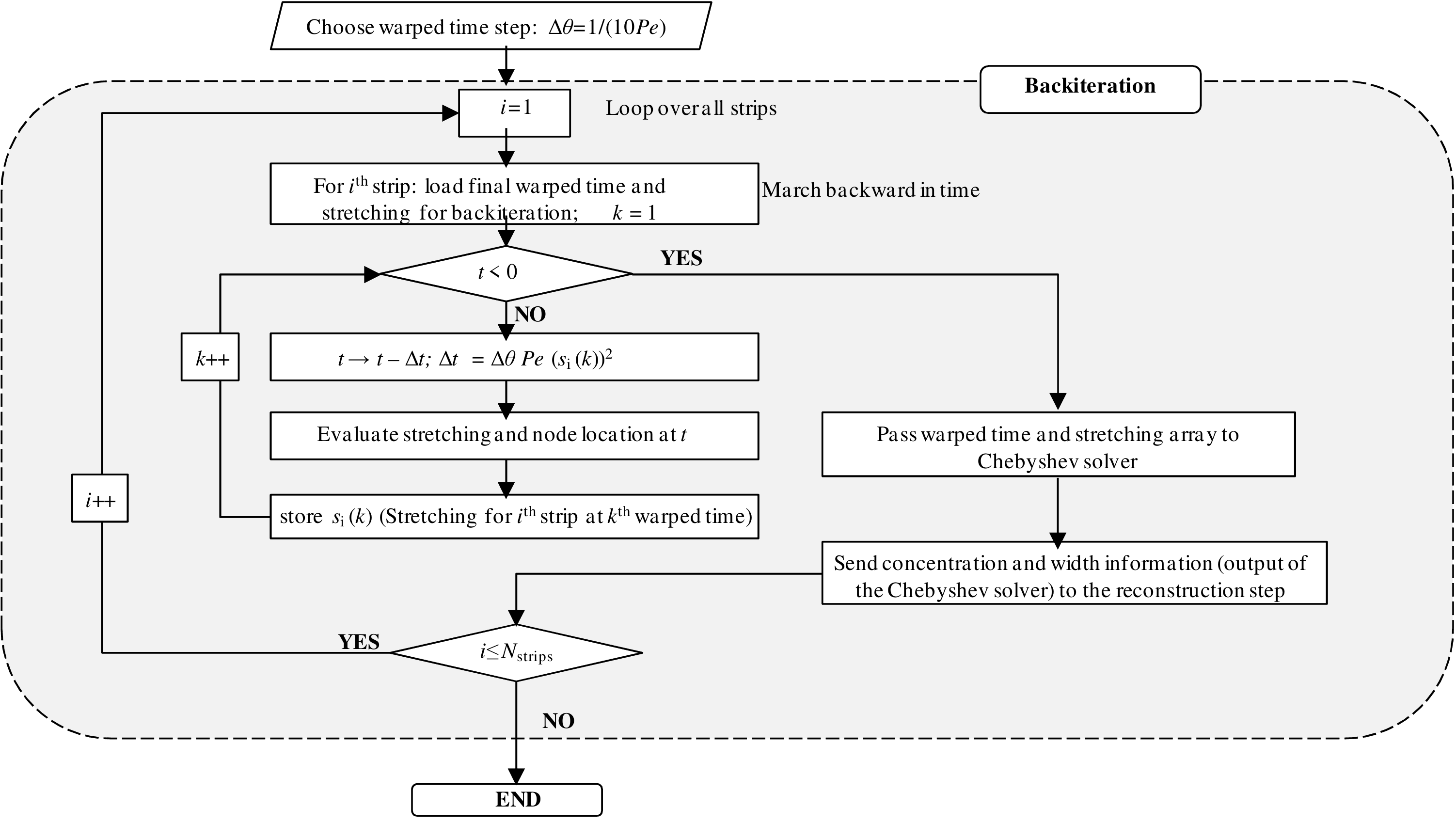}
\caption{Flowchart for the algorithm for the backiteration step of the Reactive Strip Method. }
\label{Fig:flowchart2}
\end{centering}
\end{figure}

In the present work, dynamic regridding is achieved by checking the internodal distance regular time steps (every $N_\text{update}$ steps). This is seen from the triggering of the strip refinement step in the flowchart depicted in figure \ref{Fig:flowchart1}. It must be noted here that the flowchart depicted in figure \ref{Fig:flowchart2} is a logical extension of the steps to be performed after having done the steps in the flowchart depicted in figure \ref{Fig:flowchart1}. If the internodal distance exceeds a set tolerance we insert more points in between the two nodes under consideration. For example if at the end of $N_\text{update} = 50$ steps the distance between two nodes is $l_i$ and if the tolerance for the refinement length is $l_{tol}$ then the total number of points to be inserted in between the $i-th$ and $i+1-th$ node is obtained by $N_\text{insert} = \text{ceil}(l_i/l_{tol})-1$. The separation between the nodes is found out by linearly interpolating the $N_\text{insert}$ points between the $i-th$ and $i+1-th$ node. It must be kept in mind that despite the insertions of nodes between a pair of points, the old indices must be kept the same and the number of nodes must not be updated at this stage. This ensures that in one loop over all the particles, we do not make the same refinement checks to the newly inserted nodes. It may be noted that for rapidly time-varying flows, the check-frequency for dynamically inserting new points has to be increased while also noting that the time-step for time marching should be kept reasonably small. 
Despite this, achieving the same results for a flow with high gradients using a 2D methodology would be prohibitively computationally expensive. At each nodal refinement step, the properties inherited by the new strip (the strip is defined by the pair of consecutive points) are the same strip thickness and the warped time. As we have seen earlier (equation \eqref{eq:warped}), the warped time only depends on the strip thickness, $s_i$, and is therefore constant even if a strip is broken up into several strips (nodal refinement)

The insertion of points occurs at a fixed number of specified time steps. As a result, the regions of larger stretching which transition into the lower stretching regions are subjected to a discrete discontinuous change in the length of the strips. This obviously leads to a discrete discontinuity in the warped time as well. Although we note that as the tolerance of the lengths decreases the discontinuity reduces significantly, this, however, leads to some fluctuations in the warped time in the high stretching regions of a flow. The way to obviate this issue is to reinterpolate the entire strip uniformly so that the inserted points in the strip are equally distributed over the span of the first and last nodes (the anchor nodes). Towards this, we first evaluate the local distance coordinate along the strip with respect to the first node. With this, we interpolate all the inner $x$ and $y$ coordinates along with the warped time, $\theta$, and the compression, $s$. It must be borne in mind that this reinterpolation step is not needed if the tolerance of length, the threshold length below which points must be inserted into the strip, is kept low. When the flow field is highly heterogeneous, it is best not to choose high length tolerances. In such cases the reinterpolation does not cause much difference to the otherwise default case of no interpolation. The reinterpolation step may be employed in less severe flow as it helps in achieving a reduction in run time by employing higher length tolerances to achieve the same warped time distribution. 

In figure \ref{Fig:interpolationerror} we depict the variation of the warped time for a point vortex flow as a function of the radial distance of the element from the eye of the vortex. For a point vortex, the flow field is purely tangential at any point of the field and is given by $u_{\phi} = 1/r$ where $u_{\phi}$ represents the tangential velocity. For this kind of flow it may be easily shown that the warped time is given by $\theta = \frac{t+4t^3}{3r^4}\frac{1}{Pes_0^2}$ \cite{Meunier2003}. It may be  seen that for the case where there is no interpolation, there is a fluctuation of the warped time around the analytical solution. The fluctuation is significantly reduced after the reinterpolation procedure.

The advection and dynamic refinement of strips is achieved rather easily if the velocity field is known analytically. In such a case, the particle at a position $(x(t),y(t))$ is being advected by the velocity field $(u(x(t),y(t),t),v(x(t),y(t),t))$. In an analogous manner, for the situation where the flow field is only known through a prior numerical simulation, experimental data or satellite data, the velocity field must be interpolated to the desired location at which the particle resides. In such a case, however, as the advection process is inherently unsteady in the Lagrangian frame the velocity field must be reinterpolated at every timestep at the present location of the advected nodes.

\subsection{Backiteration}

From the output of the previous step, we thus have, for a given strip, the information of all the warped times (which was regularly distributed by means of a warped time step $\Delta \theta$) and the stretching at those particular warped times. This information is critical towards solving for the concentration of the species in the local coordinate (please see the set of equations \eqref{eq:ranz}).

It was noted by Meunier and Villermaux \cite{meunierDSM} that a faithful representation of the conservative mixing process at any physical time is described by means of the warped time (which depends on the thickness of the strip of the species and on the diffusion coefficient). They conclude that at any part of the strip the maximum concentration and width are dependent on the warped time, $\theta$, as being proportional to $1/\sqrt{1+4\theta}$ and $\sqrt{1+4\theta}$ respectively. The underlying derivation for this is provided in the \ref{sec:Appendix_A}. 

The underlying assumptions towards deriving these are (a) an individual strip is independent of the dynamics of all other strips (b) the extent of the domain is infinite in the directions perpendicular to the strip and (c) diffusion in the direction of the strip is neglected. 
Assumption (a) essentially encompasses the fact that the methodology does not faithfully represent the situation near the cusps and that the method is essentially 1D. We shall also work under the ambit of these assumptions. One of the important consequences of incorporating reactions is that the maximum concentration or the width is not a well known function of the warped time. It is only in the special case of $Da \rightarrow 0$ that we may utilize the equation  (\eqref{eq:appendix_3})for determining the maximum concentration and width. It is therefore imperative to know the entire history of stretching as opposed to just the knowledge of the final warped time. For these systems, subjected to the assumptions mentioned above, it is only logical to assume that the maximum concentration occurring in the system is constant. However, because of the dependence of the system on the history of stretching, we may have, for reactive systems, a marked difference in the profiles for the same physical time. 

For the Reactive Strip Method, with the location of the strips and warped time at time $t_\text{end}$, we may now proceed to calculate the history of $s_i(t)$, which is vital information for the reaction-diffusion equations (equations \eqref{eq:ranz}) in the warped coordinates. 
We must state here that the process of computing the time history of $s_i(t)$ during the time marching procedure \ref{sec:fwd} could have been straightforward to obtain had it not been for the dynamic insertion of points based on the length and curvature of the strip. 
Moreover, maintaining a faithful representation of the interface in the time domain and obtaining the time history of the stretching in the warped time domain are fundamentally two different steps. In the former, the time marching is done in actual physical time. While in the latter case, it is convenient to determine the time history by backiterating the individual strips by starting at the final time to march back in time to the very initial location in the warped time (not the physical time). This procedure of backiteration naturally yields, for each strip, a complete description of the time history a particular strip has encountered.

Therefore, we now have a backiteration with a timestep $\Delta\theta$ for each strip. $\Delta\theta$ is also the timestep that is utilized in solving the reaction-diffusion equation in the warped coordinates. We have separately noted that in the case of reaction between the two reactants, a warped time step of $1/(10Pe)$ is more than sufficient to obtain a good convergence for the local pseudospectral Chebyshev method employed for obtaining the species concentration (section \ref{sec:cheby}).  When $Pe$ is increased, there is a decrease in the warped time. However, when the time step is also $Pe$ dependent, we obtain that the number of steps to march to the final warped time is constant. This implies that the method is independent on the $Pe$ chosen. We update the position of the strip beginning from the time $t_\text{end}$ as $x \rightarrow x - u(x,y,t) Pe s^2 \Delta\theta$ and $y \rightarrow v(x,y,t) Pe s^2 \Delta\theta$. The time corresponding to this warped time may be trivially found out by noting $t\rightarrow t - Pe s_i^2 \Delta\theta$. The new location of the points necessarily satisfies the condition of curvature and length of strip. The new warped time is found out at this warped time from equation \eqref{eq:kinematic}. We thus have the time history of the deformation $s_i(\theta)$.

\subsection{1D Chebyshev spectral collocation}
\label{sec:cheby}

\begin{figure}
\begin{centering}
\includegraphics[scale=0.7]{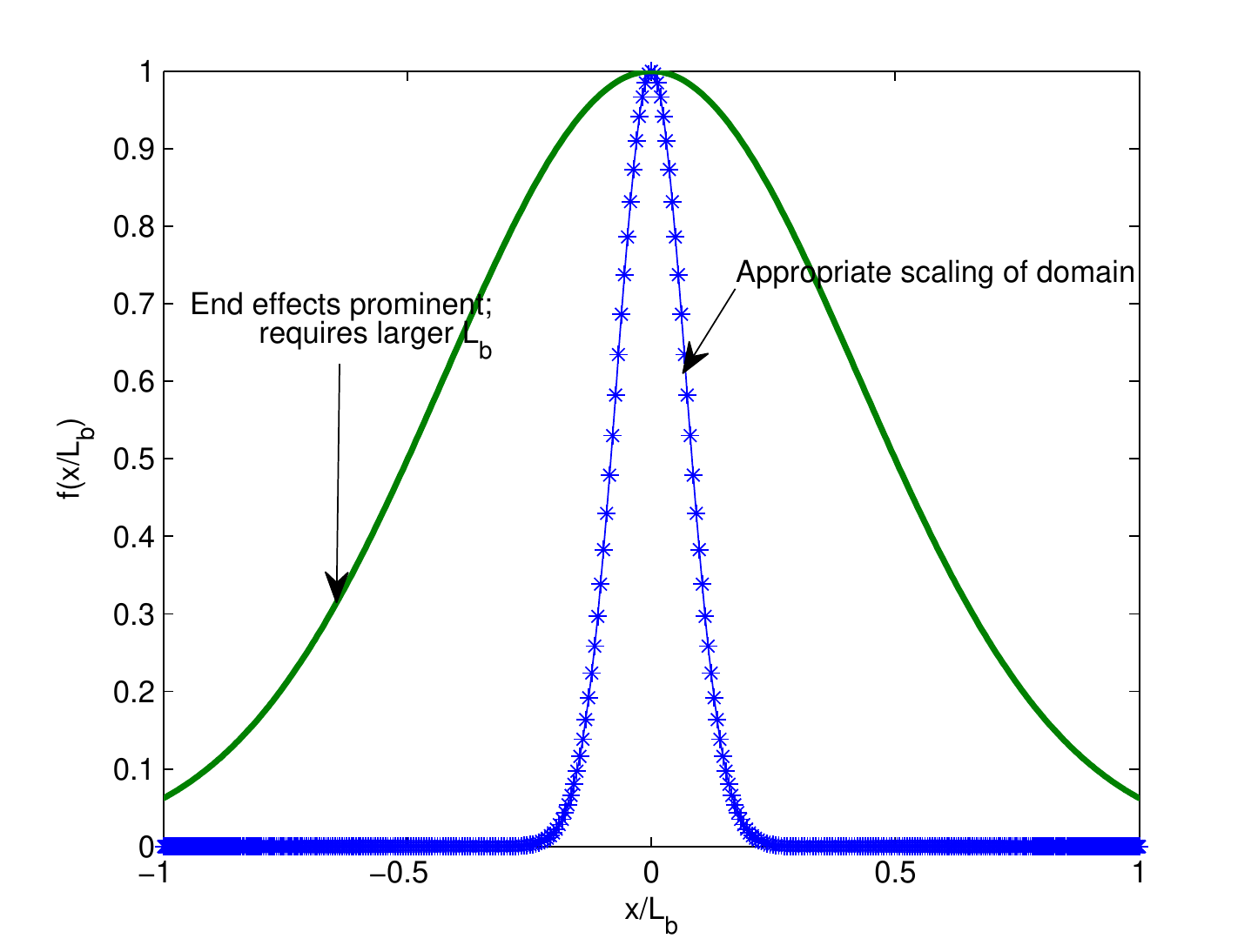}
\caption{Two curves representative of the appropriateness of the 1D domain for the spectral method. The solid line (broad Gaussian) is clearly not reaching 0 at the ends of the domain thus highlighting the end effects in its representation while the asterisk marker (narrow Gaussian) represents the scenario where the domain has been (safely) rescaled with a much larger $L_b$. The broad Gaussian is representative of a diffusion/spreading process for low P{\'e}clet number flows and large simulation times while the narrow Gaussian is representative of the diffusion/spreading process for high P{\'e}clet number flows and small simulation times.}
\label{Fig:endeffects}
\end{centering}
\end{figure}
We now use the information obtained through the previous step in order to solve the set of equations \eqref{eq:ranz}. The Chebyshev spectral collocation method provides an efficient methodology to compute the solutions in terms of linear combinations of Chebyshev polynomials. Being global methods with the basis set being Chebyshev polynomials on a Gauss-Lobatto grid (as opposed to local methods for the classical finite difference method with a basis of simple polynomials), they do not require as many grid points for the same level of accuracy. We refer to the text by Trefethen \cite{Trefethen2000} for all sorts of resources pertaining to spectral methods.

In short, we first rescale the domain so that the domain is bounded to $z=n/L_b=[-1,1]$ where $L_b$ represents the the finite extent of the 1D domain to effectively replicate an infinite domain problem. $L_b$ is usually chosen to be sufficiently large to prevent any end effects (please refer to figure \ref{Fig:endeffects}). For problems involving (a) high P{\'e}clet numbers and/or (b) early time analysis, even a moderate value of $L_b \sim O(10s_i)$ is sufficient to tackle the 1D problem. In the other case of (a) low P{\'e}clet numbers and/or (b) long time analysis, a much larger value of $L_b$ must be chosen so as to avoid the end-effects. The motivation for the choice of $L_b$ may also be made based on the distribution of the warped times as follows. It is known that the width, for a conservative scalar, is given by $s_i\sqrt{1+4\theta}$. As was earlier mentioned that in the Ranz framework, the nondimensional length is given by $z=n/s$ (please refer to equation \eqref{eq:rescaledz}). Thus, in the Ranz framework, the domain size at the final warped time has to be at least of the order of $L_b\sim\sqrt{1+4\theta}$. Therefore, the rescaling length for the Chebyshev method is chosen to be greater than  $3L_b$. 

Proceeding, let us represent the rescaled domain by $z$ and the discretized domain as $\hat{z}$, which spans over $[-1,1]$. \textit{Gauss-Lobatto} grids are employed for these kinds of discretizations which employ Chebyshev polynomials as the basis set. The discrete grid points are represented as  ${\hat z_i} = \cos \left( i\pi/N\right)$ where $N$ represents the number of elements into which the domain is discretized. 
We make use of the Chebyshev differentiation matrices to represent the spatial derivatives. The spatial derivatives are represented by means of the Chebyshev differentiation matrices multiplied to the vector representation of the pertinent scalar at the different grid points. Therefore if $\mathcal{D}$ denotes the differentiation matrix then $(\mathcal{D}\;a)$ represents the first spatial derivative of the vector $a$. $(\mathcal{D}^2\;a)$ represents the second derivative of the discrete vector for $a$ and so on. 

In a discretized notation (with the superscript $k$ denoting time index), the equation for the species $a$ may be represented as 
\begin{equation}
\frac{{{a_i^{k + 1}} - {a_i^k}}}
{{\Delta \theta }} = {\left[ {{a_i,_{zz}}} \right]^{k + 1}} - Da\,{a_i^k}{b_i^k}{\left( {{s_i ^k}} \right)^2}
\label{eq:discrete1}
\end{equation}
The time-discrete equations for the evolution of species $b$ and $c$ may be written on similar lines.
We may write the discretized form of the reaction-diffusion equations as
\begin{eqnarray}
\begin{gathered}
  a_i^{k + 1} + \Delta \theta \left[ {{a_{zz}}} \right]_i^{k + 1} = a_i^k - Da\,a_i^kb_i^k (s_i^k)^2 \hfill \\
  b_i^{k + 1} + \Delta \theta \left[ {{b_{zz}}} \right]_i^{k + 1} = b_i^k - Da\,a_i^kb_i^k (s_i^k)^2 \hfill \\
  c_i^{k + 1} + \Delta \theta \left[ {{c_{zz}}} \right]_i^{k + 1} = c_i^k + Da\,a_i^kb_i^k (s_i^k)^2 \hfill \\ 
\end{gathered} 
\label{eq:alleqs}
\end{eqnarray}
where the subscript $i$ represents the value of the variable at the $i-th$ node while the superscript $k$ denotes the $k-th$ time step. The subscript $zz$ indicates the second spatial derivative. The initial and boundary conditions for equation \eqref{eq:alleqs} may be written as
\begin{equation}
\begin{gathered}
  a_i^0 = 1\,{\text{ }}\forall x > 0,b_i^0 = 1 - a_i^0,c_i^0 = 0 \hfill \\
  a_N^k = 1,\,a_0^k = 0,\,b_N^k = 1,\,b_0^k = 0,c_N^k = 0,\,c_0^k = 0 \hfill  \text{~ .} 
\end{gathered} 
\label{eq:bcsdisc}
\end{equation}
The boundary conditions of the discretized domains are incorporated by altering the first and last rows respectively of the pertinent differentiation matrix.

\subsection{Reconstruction of concentration field}

\label{sec:reconstruct}

\begin{figure}[!th]
\begin{centering}
\includegraphics[scale=0.5]{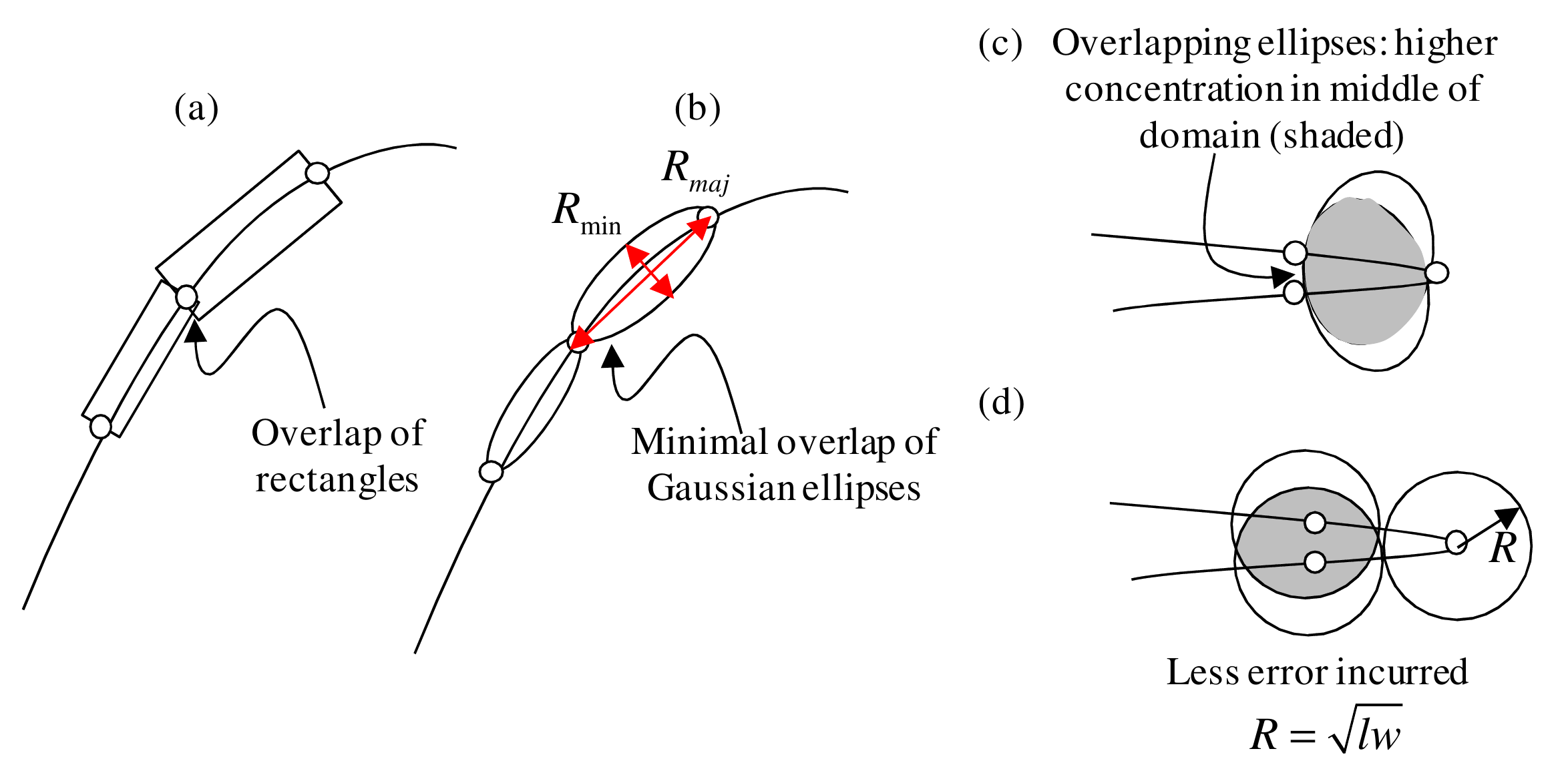}
\caption{(a) Reconstruction of the concentration field by means of Gaussian rectangular strips of length equal to the internodal distance and width obtained from an exponential distribution normal to the strip depending on the width of the concentration field obtained from the 1D spectral method. (b) Reconstruction of the concentration field by means of Gaussian ellipses with the major axis being the length of the strip and the minor axis being the width of the concentration. (c) Failure of the method in the regions of high curvature with the insertion of Gaussian ellipses. (d) The treatment of the regions of high curvature is done by inserting Gaussian circles at the nodes instead of inserting Gaussian ellipses at the midpoint of the strips. The radius of the Gaussian circle is determined by taking a harmonic mean of the major and minor axis. It must be noted that the methods illustrated in figures (c) and (d) are only to improve the reconstruction in the regions where the strip method is destined to fail. It is in these regions that the representation, in reality, may not be done by means of insertion of circles or ellipses -- both arising out of the assumption that the distribution is perpendicular to the strip; rather, these regions should be, ideally, described by means of a full 2D concentration reconstruction.}
\label{Fig:reconstruct}
\end{centering}
\end{figure}

\begin{figure}[!th]
\begin{centering}
\includegraphics[scale=0.5]{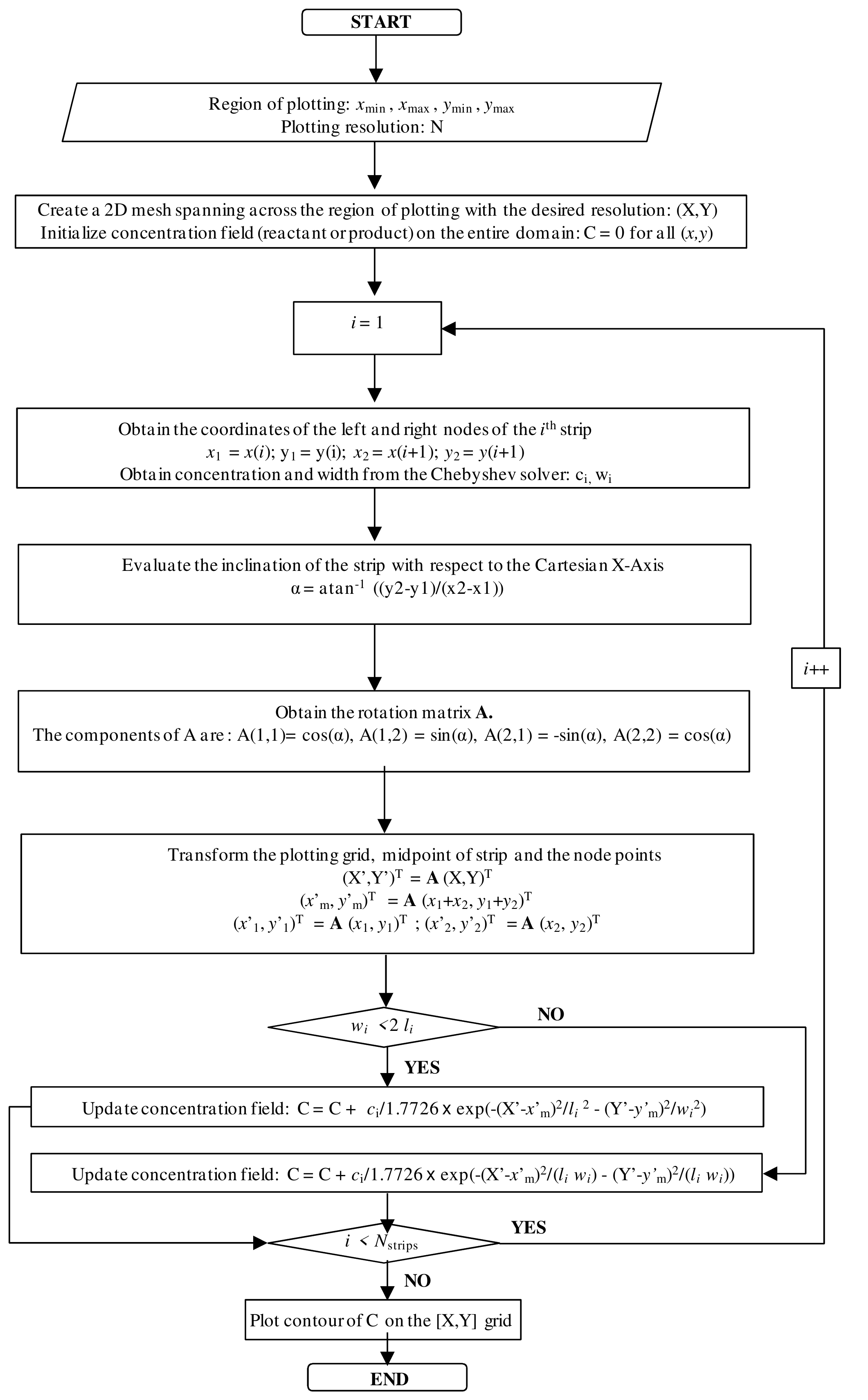}
\caption{Flowchart of the algorithm describing the reconstruction of the concentration fields based on the outputs (concentration and width of the concentration distribution) from the Chebyshev solver.}
\label{Fig:algo_reconstruction}
\end{centering}
\end{figure}

With the aforementioned methodology one obtains the solution of the 1D diffusion-reaction problem in the direction perpendicular to the strip -- we are primarily interested in the concentration distribution; quantified simply by the maximum concentration and the width (the variance of the distribution). Thus, the information may allow one to easily obtain the maximum concentration of the reactants and product along the material line. By considering a normalized  second spatial moment of the concentration distributions one may also infer the widths of these regions. As a result, one can, in practice, reconstruct the concentration field with this information in hand. Towards this goal, we  follow the methodology employed by Meunier and Villermaux \cite{meunierDSM}. The idea for reconstruction is essentially the same because it does not rely on the mechanism of production or the time history. The method for reconstruction merely utilizes the information about the concentration at a material line and the width of the concentration distribution about that particular material line. We mention the procedure for reconstruction we have adopted in this work for completeness.

Along each strip we may place a rectangular concentration field which decays perpendicular to the strip based on the 1D solution. The length of the rectangular concentration field is, in that case, equal to the length of the strip. The main issue in utilizing this method is that the concentration field near the nodes suffers an error in representation even at relatively low curvatures (please see figure \ref{Fig:reconstruct}(a). To avoid this problem instead of having sharp edged rectangles, we utilize Gaussian ellipses instead which are inserted with the center of the ellipse placed at the midpoint of the two nodes. The major axis of the Gaussian ellipse is chosen as the internodal length (length of the strip) while the minor axis is chosen in accordance to the width of the concentration determined through the 1D solution. This method is elucidated in figure \ref{Fig:reconstruct}(b). We have also depicted the flowchart of the algorithm for reconstructing the concentration field in figure \ref{Fig:algo_reconstruction}.

The reconstruction may be done in the following way. We define the grid over which the reconstruction is desired -- let us denote it by $[X,Y]$. This may be the entire extent of the strip or may just be a zone of interest. Over this domain we need to find out the orientation of a strip with respect to the horizontal direction (this is easily evaluated by means of $\phi = \tan^{-1}(\Delta y/\Delta x)$ where $\Delta y$ and $\Delta x$ represent the difference in the $x$ and $y$ coordinates respectively (this must be obtained by the use of the MATLAB function {atan2})). The rotation matrix to \textit{transform} this regular Cartesian mesh so that the major axis is aligned with the strip is given by $A_{1,1} = \cos(\phi)$, $A_{2,1} = \sin(\phi)$, $A_{1,2} = -A_{2,1}$ and $A_{2,2} = A_{1,1}$. The Cartesian mesh in the rotated domain is denoted by $[X', Y']$. For this case we now have the rotated coordinates as simply ${\bf{x}} \rightarrow {\bf{A}}{\bf{x}}$. Now, for each strip we know (a) the length of the strip, $l_i$ and (b) the concentration width $w_a$ (or $w_c$) from the 1D spectral method.  With this, the concentration profile in this frame is simply obtained as 
\begin{equation}
c_i = \frac{a_i}{1.7726}\times \exp \left( -\frac{(X'-x_m)^2}{l_i^2} - \frac{(Y'-y_m)^2}{w_{i,a}^2}\right),
\label{eq:conc_reconst}
\end{equation}
where the factor 1.7726 originates from the overprediction in the concentration at the midpoint due to the contribution of all the neighboring Gaussian contributions \cite{meunierDSM}. Therefore, the concentration reconstruction is achieved by plotting the above obtained concentration ellipse in the $[X,Y]$ domain.

In the case where the width of the concentration of a given species is larger than the length of the strip (which is actually a violation of the assumption that the characteristic length scale in the axial direction be thinner than that in the perpendicular direction) the reconstruction is not in tune with the physical reality. Before delving into the corrective measures of representing such elements we must point out that in most flows such strips are only a very small fraction of the total number of strips. As such, even if strips are not carefully resolved the overall reconstruction is only slightly affected. Nevertheless, one of the more appropriate resolution of this issue, as performed by Meunier and Villermaux \cite{meunierDSM}, consists in redistributing the warped time and inserting Gaussian circles at the nodes instead of the midpoints. Accordingly, the radius of the Gaussian circle is chosen to be the geometric mean of the, otherwise evaluated, length and width of the strip.

\section{Results and discussion} \label{sec:result}
To demonstrate the methodology presented in this work we consider three different prototypical flows \cite{meunierDSM} which are quite relevant for many other naturally occurring flows as highlighted in the Introduction. 
We first consider the case of a linear shear flow in which the stretching is expected to be linear in time, which is the first order approximation of flow fields where the flow curvature is relatively small \citep{Leal}. 
Besides this, we also apply the analysis presented here for the case of a point vortex \cite{Meunier2003}. Such a flow field is important to model unstable interfacial shearing processes (Kelvin-Helmholtz instability) between two reactive fluid layers \cite{emilio} having markedly different densities.  The choice of the flow field has also been motivated by the observation that maximum mixing occurs in the eddies and vortical structures formed during the flow. 
Finally we present the present methodology for the case of a chaotic sine flow \cite{meunierDSM}. The chaotic sine flow presents the underlying flow field to model typical stirring processes in several industrial process. The results for an unreactive species are compared against analytical solutions of the warped time and concentration fields for the cases of a linear shear and point vortex. We then focus on depiction of the concentration distributions for reactive species for the three aforementioned flows. 

\subsection{Linear shear flow} \label{sec:linflow}

The simplest case under consideration is the linear shear flow. This prototypical flow is widely studied as it offers a relatively simple advective component \cite{emilio}. 
Such a flow is also physically relevant from the point of view of a local analysis wherein the locally linearized flow field may be viewed, at leading order, as a linear shear flow \citep{Leal}. Because of the uniformity of the linear shear, the distribution of the warped time is uniform over all the different strips. 
Essentially this scenario is reminiscent of the case where the entire lamella behaves as a solitary strip. 
It may be noted that for a linear unit shear flow  the thickness of the lamella varies as $s = 1/\sqrt{1+t^2}$. In this case the warped time may be easily obtained by integrating equation \eqref{eq:warped} for times larger than 1 to yield $\theta = \frac{t^3}{3Pe}$. The width of a conservative species evolves as a function of the warped time as $\sqrt{1+4\theta}/\sqrt{1+t^2}$ which, in the limit of $t\gg 1$, yields a width which is proportional to $\frac{\sqrt{t}}{\sqrt{Pe}}$. This result essentially underpins the scaling which dictates that the boundary layer becomes thinner as $1/\sqrt{Pe}$ while also indicating that the width goes as $\sqrt{t}$.

We mention here a note about the efficiency of the method presented in this work. We consider the test case of a linear shear flow with $Pe = 10^3 and 2\times 10^3$ and $Da = 1$. The runtime of the methodology proposed here does not depend on $Pe$, as we had noted earlier. For a single core implementation on MATLAB with a i7-4790 CPU (3.6 GHz), it took 3 minutes and 2 seconds with $\Delta t = 10^{-2}$ and $l_\text{tol} = 0.025$. However, for simulating the same system with COMSOL, it took, on a 4 core implementation of the same machine, 23 minutes and 5 seconds with a mesh size of 0.025 and $\Delta t = 10^{-2}$ for $Pe = 10^3$. The same system with $Pe = 2\times 10^3$ took 40 minutes and 10 seconds to run with the same simulation parameters as above. Clearly, the FEM implementation starts becoming slower as the $Pe$ is increased while the RSM remains independent of $Pe$.
\begin{figure}
\begin{centering}
\includegraphics[scale=0.7]{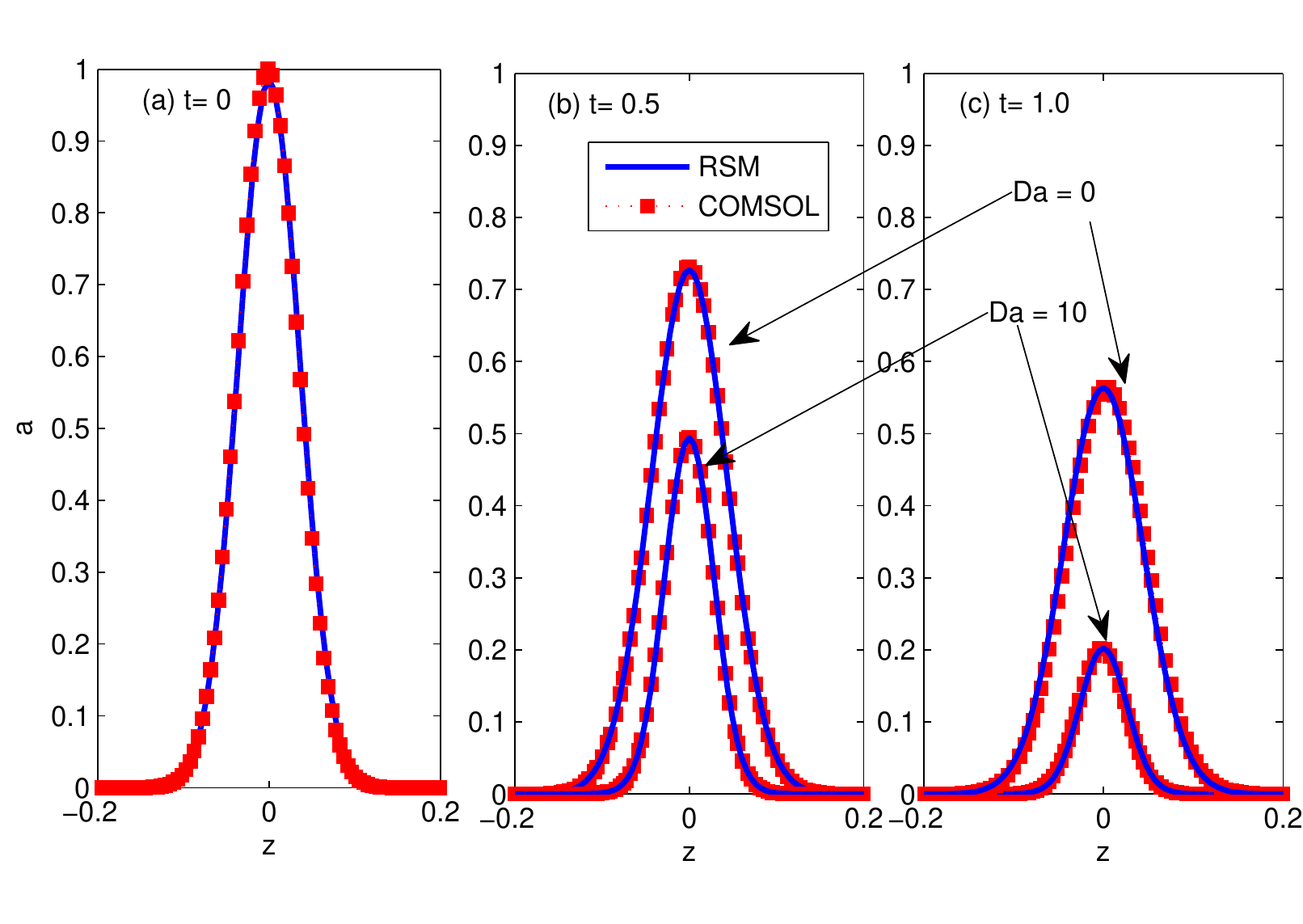}
\caption{Concentration profiles of the species $a$ residing in the lamella for a shear flow at $t = $ (a) 0, (b) 0.5 and (c) 1.0 for $Pe = 1000$. The results obtained by means of the present reactive strip method is denoted by the legend \textit{RSM} while that obtained from the full 2D FEM based COMSOL simulation is denoted by the legend \textit{COMSOL}. The profiles are plotted along the line which is perpendicular to the direction of the strip at that particular instant of time. The effect of $Da_\text{II}$ has also been highlighted in the figures at $t = 0.5, 1.0$.  }
\label{Fig:sf_a_comsol_compare}
\includegraphics[scale=0.7]{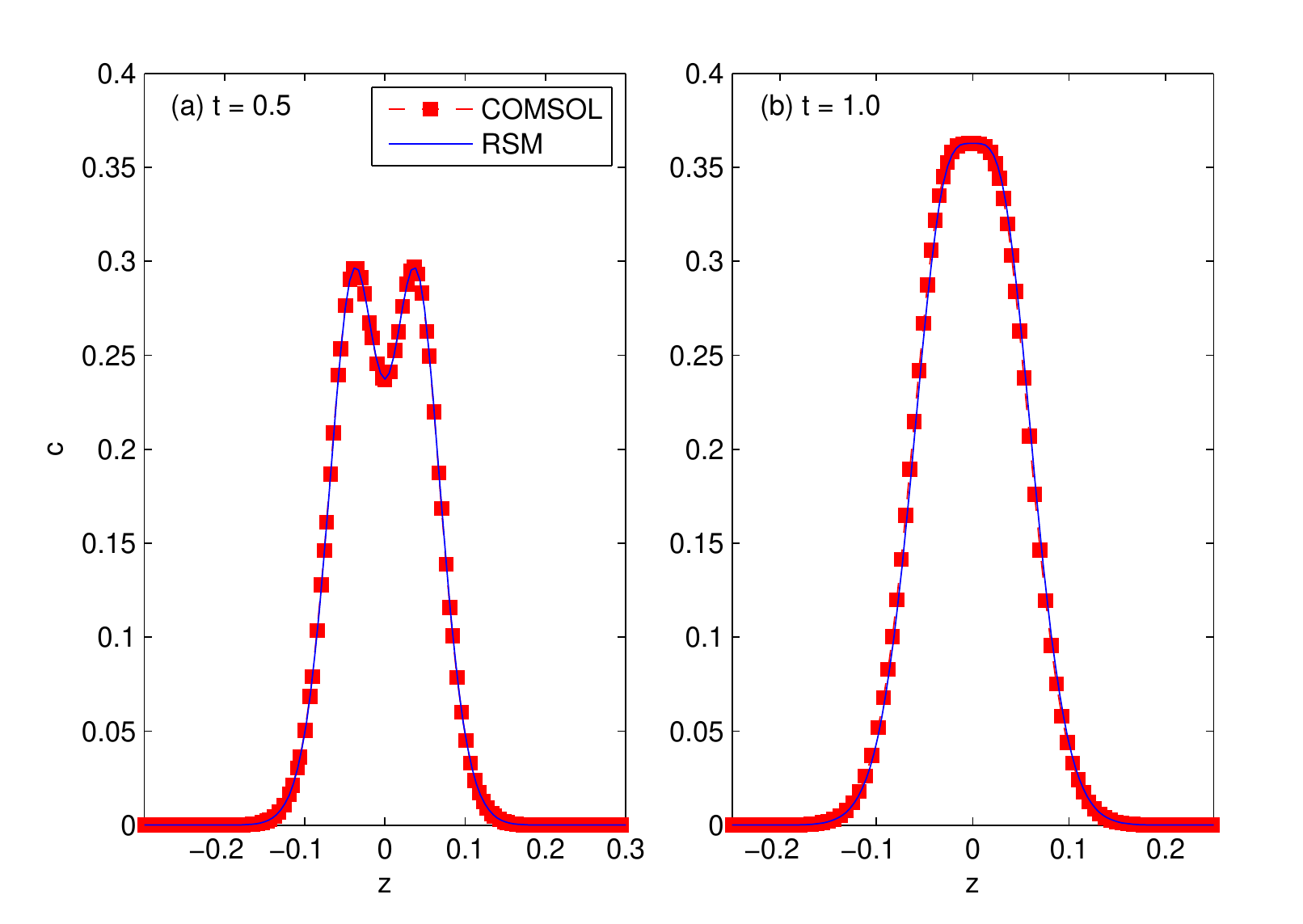}
\caption{Concentration profiles of the species $c$ for a shear flow at $t = $ (a) 0.5 and (b) 1.0 for $Da_\text{II} = 10$ and $Pe = 1000$.}
\label{Fig:sf_c_comsol_compare}
\end{centering}
\end{figure}

In figure \ref{Fig:sf_a_comsol_compare} we depict the concentration profile of the reactant residing in the lamella for the case of a linear unit shear flow at times $t = $ (a) 0, (b) 0.5 and (c) 1.0 for $Pe = 1000$. The solid lines depict the curves obtained from the reactive strip method while the symbols represent the curves obtained from the 2D COMSOL simulations. In the figures we have also depicted the effect of the Damk\"ohler number ($Da_\text{II}$ = 0, 10) on the concentration profiles obtained through the two aforementioned methods. The concentration profiles are  perpendicular to the direction of the material line which is subjected to the shear flow. In order to achieve this in the 2D simulation we perform cross sectional plots at an angle of $\arctan(-t)$. It may be seen right away that the reactive strip method yields excellent agreement with the 2D simulations. The agreement of the concentration of the reactant is also maintained at reasonably high $Da_\text{II}$. 

In figure \ref{Fig:sf_c_comsol_compare} we depict the concentration of the product formed at the interface of the two reactants for the case where $Da_\text{II} = 10$ and $Pe = 1000$ at times $t = $ (a) 0.5 and (b) 1.0. At early times, since the influence of the chemical kinetics is relatively high, we observe two peaks in the cross section plot. As time progresses, the two peaks coalesce and form one single peak whose maximum is located at the center of the strip, each peak corresponding to the localized rate of reactions at the interfaces of the lamella. As before, the concentration profiles obtained from the two methods are in excellent agreement with each other.  

\begin{figure}[!ht]
\includegraphics[scale=1]{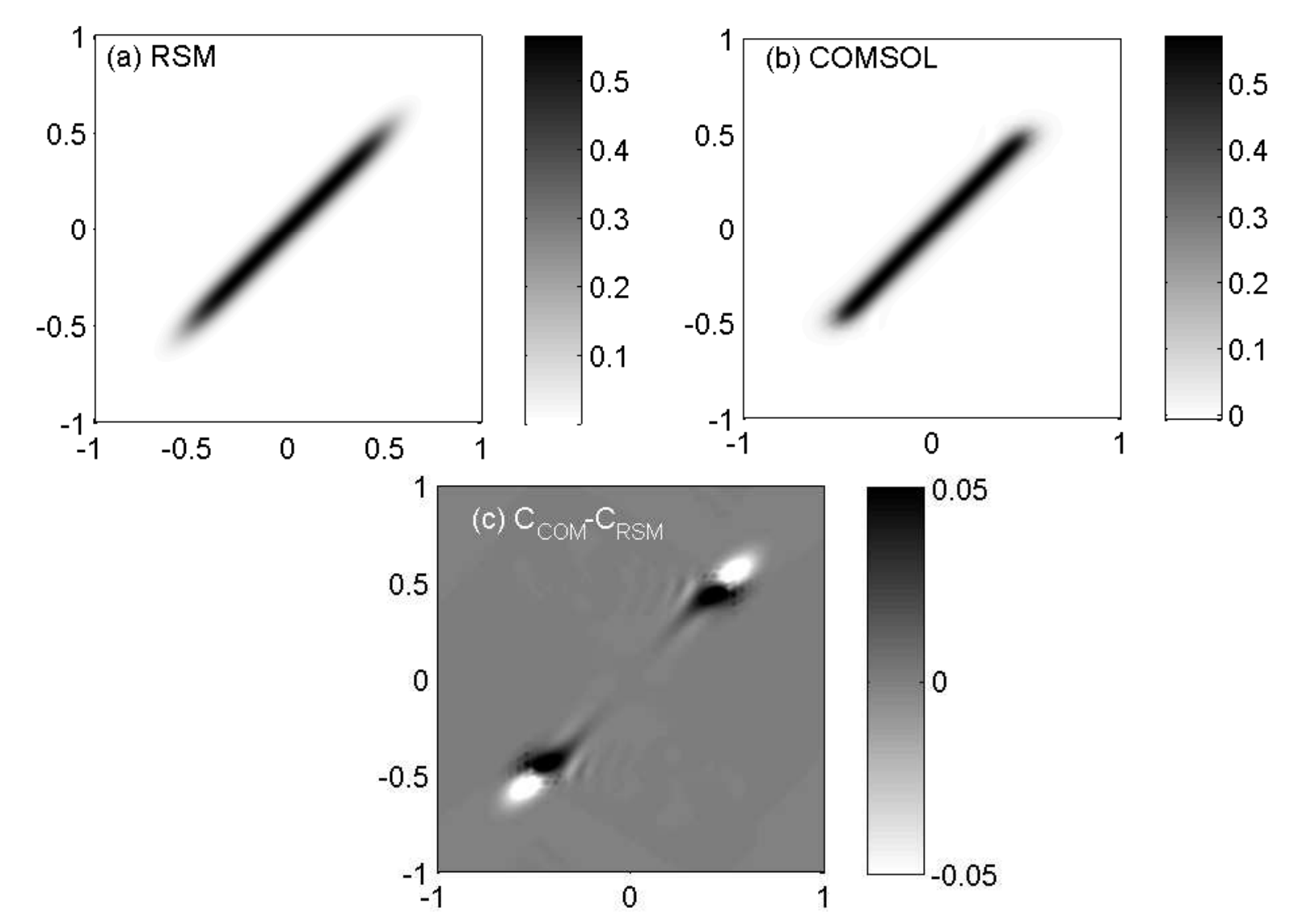}
\caption{Difference in the reconstructed concentration field (for $Pe = 1000$ and $Da_\text{II} = 0$) obtained through the description of the reconstruction methodology in section  \ref{sec:reconstruct} as shown in (a) and the concentration field found through 2D simulation in COMSOL (b). Subplot (c) depicts the difference in the concentration fields obtained in (a) and (b) do highlight the error incurred at the edges (where the assumption that the system is locally 1D loses its validity.}
\label{Fig:diff_conc_Da_0}
\end{figure}

In the case of a shear flow, it may be confirmed from the numerical estimates that the maximum concentration occurring at the centerline and the width of the species are accurately given by the analytical predictions $1/\sqrt{1+4\theta}$ and $s(t)\sqrt{1+4\theta}/\sqrt{2}$.  For example, the warped time ($\theta(t) = \frac{t+t^3/3}{Pes_0^2}$) obtained from the reactive strip method for $Pe = 10^{3}$ and $s_0 = 0.05 $ at $t = 1$ is obtained analytically as $0.53333$ while that obtained from reactive strip method is $0.5339$ which yields a relative error of $0.11\%$. It has been confirmed independently of the validation shown in figure \ref{Fig:sf_a_comsol_compare} that the widths obtained from the reactive strip method ($0.04441$) differ from the analytical estimate ($0.04424$) for $Da_\text{II} = 0$ at time $t = 1$ by a relative error of $0.4\%$.

We proceed to demonstrate the reconstruction of the concentration field from the reactive strip method as compared to the full 2D simulation obtained from COMSOL. In figure \ref{Fig:diff_conc_Da_0} we depict the difference in the reconstructed concentration profiles obtained using the reactive strip method and the 2D simulations from COMSOL. We may observe from the figure that the main source of error ($\sim 10\%$) is primarily at the edges. The error is due to the following reason. The main assumption behind the reconstruction is the fact that concentration in a strip diffuses only perpendicular to it. Consequently, this assumption loses its validity in the region near the edges of the lamella where diffusion is essentially two-dimensional. It may be noted that at high P{\'e}clet numbers, the outward axial diffusion near the edges also becomes smaller allowing for a much better approximation to the full 2D concentration field.

\subsection{Point vortex flow} 
\label{sec:vortexflow}

We now move on to the case of advection-reaction-diffusion acted upon by a point vortex flow. Contrary to the case of a fixed linear shear rate, the velocity field is radially dependent, leading to a spatially dependent rate of shear. 
The manifestation of this is in  terms of a spatially dependent warped time which depends on  $t^3 r^{-4}$ at times larger than $1$. This may be clearly seen from the expression and from the simulation results (please refer to figure \ref{Fig:interpolationerror}. 

Let us consider, for example, the experimental case of Meunier and Villermaux \cite{Meunier2003}. It may be observed that the lengthscale under consideration is chosen to be the extent of the half-domain under consideration, i.e. $\mathcal{L} = 2.4$ cm. The scale of velocity is the circulation per unit average radius and is chosen to be $U = 14.2/2.4$ cm s$^{-1}$. Based on the data for diffusivity, $D\sim 10^{-6} \rm{cm^2/s}$, the P{\'e}clet number is determined to be of the order $O(10^{5}-10^{6})$ which corresponds to a relatively large P\'eclet number..  

\begin{figure}
\begin{centering}
\includegraphics[scale=0.6]{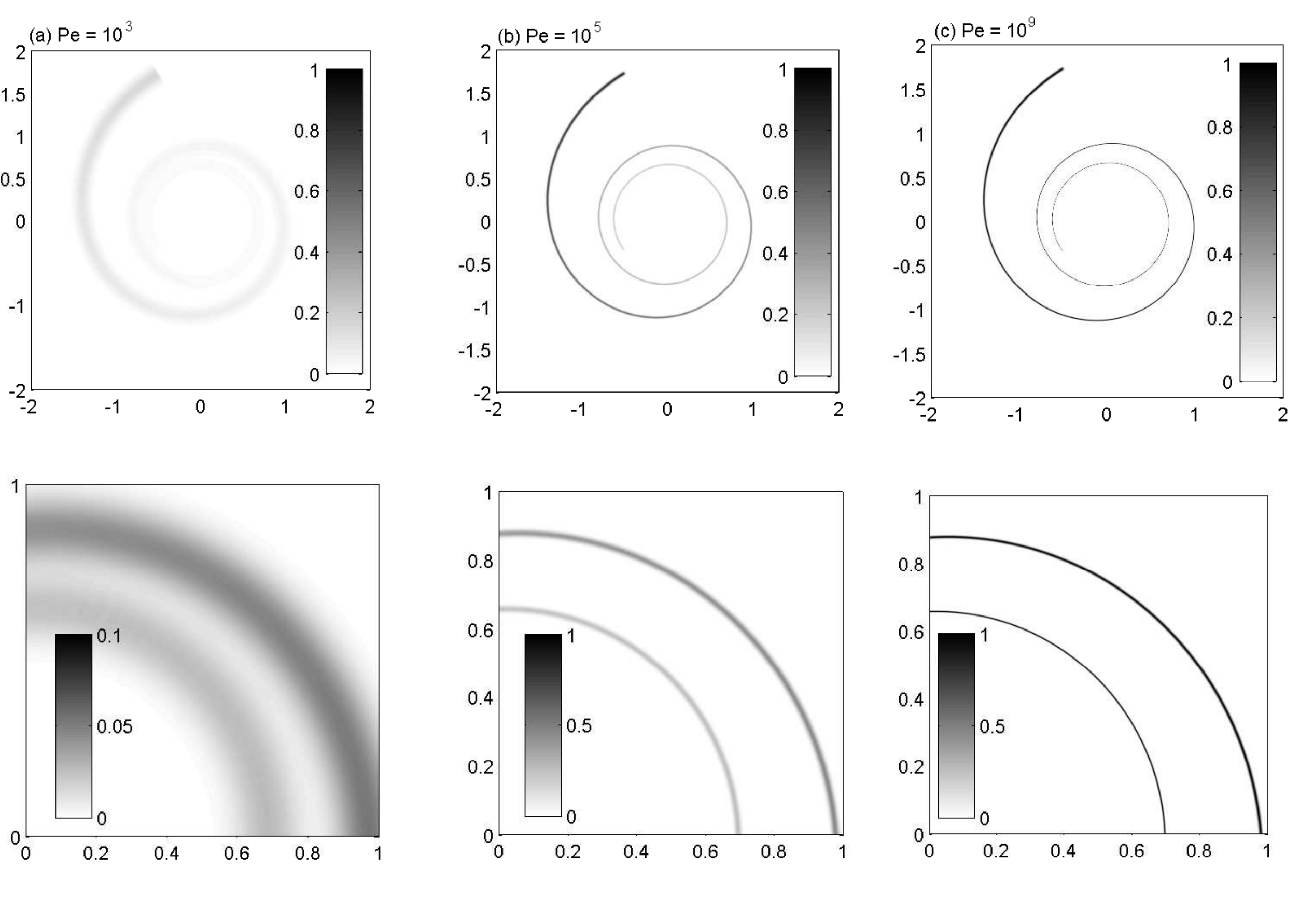}
\caption{Concentration profile of the reactant obtained for a nonreacting species ($Da_\text{II} = 0$) subjected to a point vortex flow. The three cases correspond to the concentration of the reactant at $t=6$ for (a) $Pe = 10^3$, (b) $Pe = 10^5$, and (c) $Pe = 10^9$ respectively. The lower row depicts the magnified portions of the concentrations fields depicted in the top row for the region bounded by $(x,y) \in [0,1]\times [0,1]$.}
\label{Fig:conc_profile_reactant_point_vortex}
\includegraphics[scale=0.65]{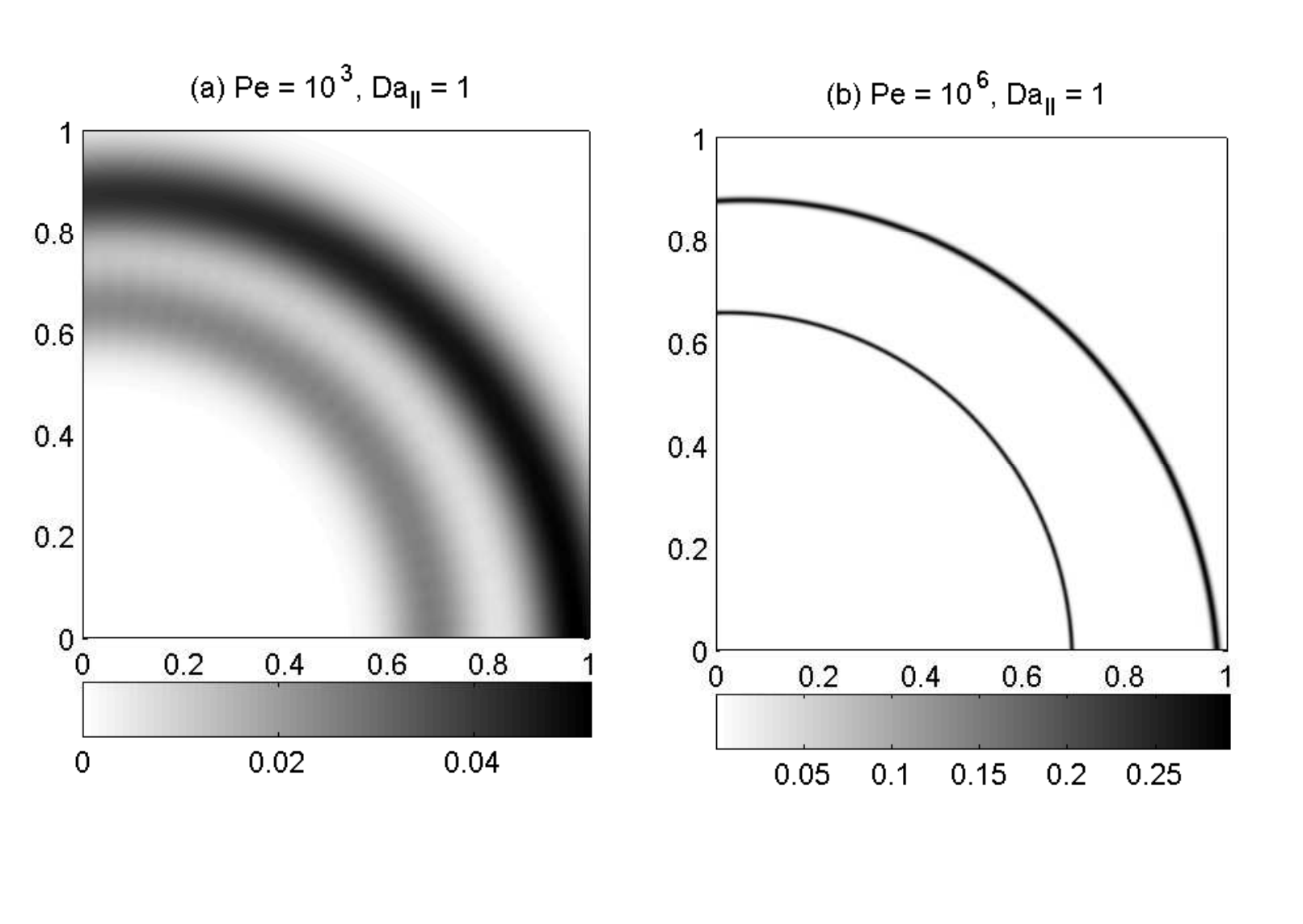}
\caption{Surface plot of the product's concentration field in a point vortex flow for $Da_\text{II} = 1$ and (a) $Pe = 10^3$ and (b) $Pe = 10^6$ at $t = 6$.}
\label{Fig:conc_c_vortex_Da_1}
\end{centering}
\end{figure}

With this physical significance in mind, in figure \ref{Fig:conc_profile_reactant_point_vortex} we depict the distribution of a conservative scalar subjected to a point vortex described in section \ref{sec:fwd} at $t = 6$ (please also refer to \cite{Meunier2003} for more details on the nature of the flow). The initial concentration field is a Gaussian defined as per $a_0 = \exp(-y^2/0.06^2)$ with a range in $x$ limited to $0.6<x<1.8$. It may be observed from the figure that as $Pe$ is increased, we obtain a significant reduction in the width of the concentration about the material line which is wrapping around the eye of the vortex. This also reflects on the fact that the method is well suited for high $Pe$ flows. Moreover, the results shown in figure \ref{Fig:conc_profile_reactant_point_vortex} b appear to be in good agreement with the experimental images of Meunier and Villermaux \cite{Meunier2003}.  Another aspect that we would like to address in terms of the validity of the methods is what occurs at the eye of the vortex. Generally, this is the region in which the neighbouring strips come so close to each other that these concentrations overlap. For two parallel strips, this consideration does not lead to any issues. This is a problem only when this occurs near a bend where the 1D approximation is not valid. The concentrations in the straight sections are additive for the case of a conservative scalar. This assumption breaks down further due to considerations of the nonlinear reaction term. The analysis is, however, still valid in the limit of low Damk\"ohler numbers. This may be  verified by writing down the set of governing equations as a regular perturbation in $Da_\text{II}$ and proceeding to isolate various orders of $Da_\text{II}$ with each order yielding a set of linear partial differential equations. We shall not dwell further on this issue here.

In figure \ref{Fig:conc_c_vortex_Da_1} we depict the surface plot for the product formed for the two different cases of (a) $Pe = 10^3$ and (b) $Pe = 10^6$ with $Da_\text{II} = 1$ at $t=6$. It may be noted that for larger $Pe$ we obtain a thinner region of the product albeit at a much higher concentration than the case of a low $Pe$. Diffusion being a dominant mechanism for systems with low P\'eclet number causes much broader widths.

\subsection{Chaotic sine flow} 
\label{sec:sinflow}

The aim of this method is to go beyond such well defined flow fields and investigate more complicated cases pertaining to \textit{chaotic} mixing. Towards this we consider a chaotic sine flow  defined as follows

\begin{align}
\left( u,y \right) &= V_0 \left( 0, \sin(2\pi x + \chi_m^{x}\right),\; m<t<m+1/2\\
\left( u,y \right) &= V_0 \left(\sin(2\pi y + \chi_m^{y}, 0\right),\; m+1/2<t<m+1,
\label{eq:c_sine}
\end{align}
where the phases, $\chi_m^{x}$ and $\chi_m^y$, of the flow are randomly chosen at each period. We have employed the same values of the phases as that employed by Meunier and Villermaux \cite{meunierDSM}. We reproduce here the various values of the phases in table \ref{Tab:tab1} for the convenience of the reader.
\begin{table}[!th]
\centering
\begin{tabular}
{c c c c c c c c}
\hline
$m$ & 0 & 1 & 2 & 3 & 4 & 5 & 6 \\
$\chi_m^x$ & 1.2154 & 4.2865 & 1.9023 &  3.4034 & 0.9480 & 4.3850 &  2.3774 \\
$\chi_m^y$ &  3.1199 & 5.6534 & 5.1624 & 4.0521 & 5.1395 & 4.1483 &  2.1487 \\
\hline
\end{tabular}
\caption{$x$ and $y$ phases of the chaotic sine flow}
\label{Tab:tab1}
\end{table}

\begin{figure}[!th]
\begin{centering}
\includegraphics[scale=0.7]{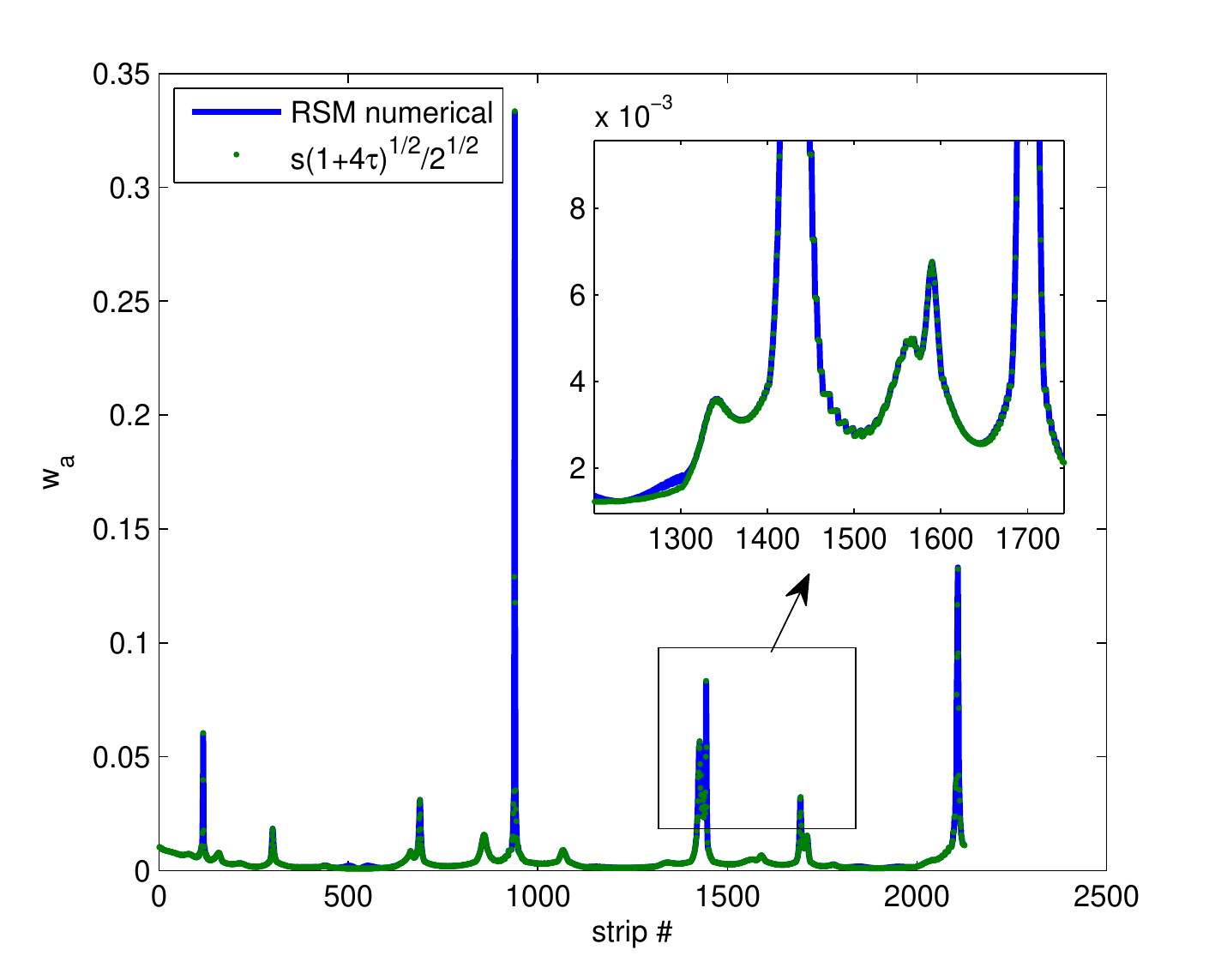}
\caption{Variation of the width of a conservative scalar at $t=6$ for an imposed sine flow as determined from the reactive strip method (solid line) compared to the analytical solution (point markers). The inset shows the magnified portion of the plot. }
\label{Fig:warped_time_chaotic_sine}
\end{centering}
\end{figure}
We first depict in figure \ref{Fig:warped_time_chaotic_sine} the variation of the width of the reactant $a$ at $t = 6$ for the chaotic sine flow depicted by \eqref{eq:c_sine} as compared to the analytical 1D prediction. The comparison entails the fact that $Da_\text{II}=0$. The 1D theory predicts the width to be given by $s\sqrt{\frac{1+4\theta}{2}}$ (please refer to the \ref{sec:Appendix_A}). The numerical predictions from the reactive strip method are seen to be in remarkable agreement with the theoretical predictions. 

\begin{figure}[!th]
\includegraphics[scale=0.6]{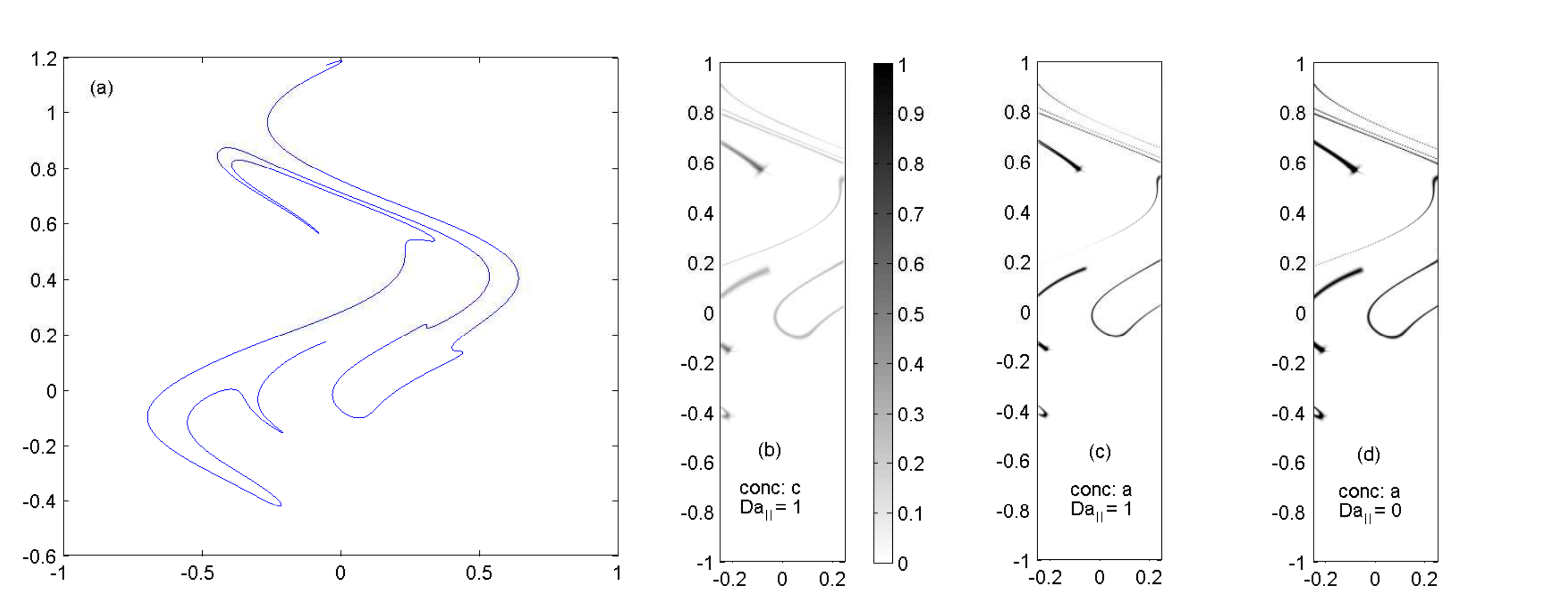}
\caption{(a) Material line at $t = 5$, after it has evolved from an initial vertical straight segment at $x = 0$ and $-0.5<y<0.5$ . (b) Concentration of product ($c$) and (c) concentration of reactant $a$  obtained at $t = 5$ for $Pe = 10^6$ and $Da_\text{II} = 1$ for the region defined $-0.25<x<0.25$ and $-1<y<1$. (d) Concentration of the reactant $a$ for the case of a conservative scalar for $Pe = 10^6$.}
\label{Fig:Chaotic_sine_surface}
\end{figure}

\begin{figure}[!th]
\begin{centering}
\includegraphics[scale=0.8]{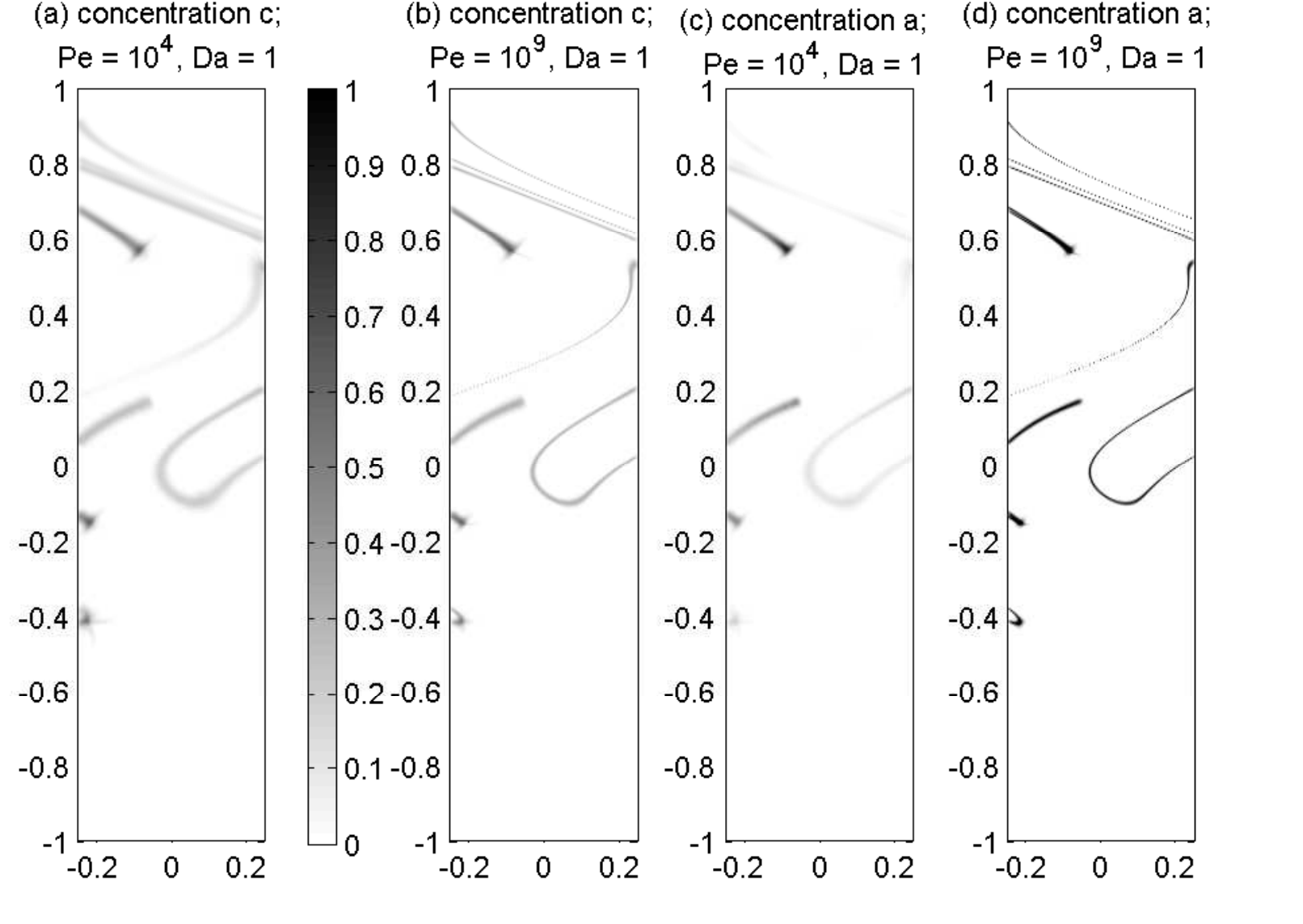}
\caption{The various concentration distributions obtained for the material line affected by a chaotic sine flow at $t = 3$ for concentration of product for (a) $Pe = 10^4, \;Da_\text{II} = 1$, (b)$Pe = 10^9,\; Da_\text{II} = 1$, and concentration of the reactant at (c) $Pe = 10^4,\; Da_\text{II} = 1$ and (d) $Pe = 10^9, \;Da_\text{II} = 1$.}
\label{Fig:chaotic_sine_Da_1}
\end{centering}
\end{figure}

In figure \ref{Fig:Chaotic_sine_surface}, we depict the nature of the concentration distribution on a material line acted upon by a chaotic sine flow (described by equation \eqref{eq:c_sine}. The material line which is initially a straight line oriented along the $y$ axis between $y = -0.5$ and $y=0.5$. In the case depicted here, the P{\'e}clet number is chosen to be $10^6$ and the Damk\"ohler number is $1$ (when defined). In subfigure (a) the  state of the material line is depicted at $t = 5$. In subfigures (b) - (d) we depict only a portion of the material line between $-0.25<x<0.25$ and $-1<y<1$ for brevity. In subfigure (b) we depict the concentration of the product formed over the material line. In subfigure (c) we depict the concentration of the reactant originally present as a Gaussian over the initial material line. In these two cases $Da_\text{II} = 1$. In subfigure (d), for comparison, we depict the concentration of the reactant $a$ for the case without any reactions, i.e. $Da_\text{II} = 0$. It may be observed from the concentration distributions that regions involving a relatively larger stretching (straight long segments) exhibit a much smaller width. The concentration of the reactant has almost vanished from this region owing to the increase in the gradient which promotes the diffusion and consequent reaction. On similar lines, it may be noted that in regions with relatively larger bends, the concentration is much higher. If we compare the concentration of the reactant against the case where there is no reaction, it may be clearly observed that the nonreactive case has a larger width at the same location of the material line. The lower width of the reactant concentration for $Da_\text{II} = 1$ is attributed to the fact that the larger spread of reactants is accordingly consumed by the other reactant to lead to the formation of the product.

In figure \ref{Fig:chaotic_sine_Da_1} we depict the concentration of the product formed and the inner reactant for the two cases of $Pe = 10^4, \;Da_\text{II} = 1$ and $Pe = 10^9, \;Da_\text{II} =1$. From the figure it is apparent that for lower $Pe$ the spread of the species is, understandably, larger which leads to lower concentration along regions of high stretching.

\section{Conclusions} \label{sec:conclusion}
	The method presented here reports here an efficient methodology for accounting for reaction in high P{\'e}clet number mixing. By tracking the time history of stretching of the elementary strips of the material line in any arbitrary flow field, we are able to accurately obtain the information about the local concentration distribution of the reactants and products about that particular strip. The methodology consists of tracking the concentration fields not in the Eulerian frame, but rather focusing on the dynamics about the Lagrangian frame attached to the different elementary strips which comprise the entire concentration distribution that we are interested in. The problem of incorporating reaction in the Lagrangian method implies that the system dynamics strongly depends on the nature of stretching, which is  not the case for a conservative scalar where the final location and warped time are enough to describe the concentration distributions. We have validated the methodology presented here for the limiting case of unreactive species which are in excellent agreement with the analytical results for the solution of diffusion of a lamella in the Lagrangian frame for a linear shear flow and point vortex flow. We have also presented results for reactive species for the cases of a linear shear flow, point vortex flow and a chaotic sine flow. 
	
	We also enumerate the limitations of the method as well. 
This methodology is not appropriate for situations in which the transport impacts the flow, as in the case for flows driven by heterogeneities in density or interfacial tension arising from changes in a scalar such as concentration or temperature.	Besides this, when the reaction term is nonlinear, the method is unable to account for the case where the strips overlap or merge together or aggregate together. We would also like to point out that the methodology is well hastened to account for individual strips of reactants immersed in another reactant such that there is no significant folding of the strip. This is true of several kinds of flow (up until the point of merging) such as combustion and other kinds of open flows. Even for the case of closed flows, the methodology described in this work would be able to describe reactive flows up until severe stretching and folding.
	
	Besides this, the methodology has to be appropriately modified to take care of the cusps and cusp-like structure that are formed in cases of complex flows. We remind the reader that these issues are chronic of this methodology as was also noted by the work by Meunier and Villermaux \cite{meunierDSM}. Extension of this method to 3D reactive flows - where the \textit{embedded} filament would be represented by a sheet is the subject of current investigation. Besides the obvious issue of dynamic triangulation, the memory requirements for chaotic 3D flows need to be appropriately handled in order for the procedure to be manageable. 
	
	\section{Acknowledgement}
	AB gratefully acknowledges postdoctoral funding through the \textit{Agence Nationale de la Recherche} (ANR-14-CE04-0003-01 subsurface mixing and reaction).

\appendix
\section{Diffusive transport of a single lamella}
\label{sec:Appendix_A}
Here, we derive the fundamental form of the velocity field which is utilized for the reduction of the 2D governing equations to 1D. We also obtain the solution for a single diffusing lamella. 

In the vicinity of a node $i$, the velocity in the local coordinate system, $(n,\sigma)$, can be written as 
\begin{equation}
v_{\sigma} = - \frac{n}{s_i} \frac{d s_i}{dt} + \frac{\partial v_{\sigma}}{\partial n} {n};
\;\;v_{n} = \frac{n}{s_i} \frac{ds_i}{dt},
\end{equation}
where $v_\sigma$ denotes the velocity  component along the local streamline and $v_n$ denotes the velocity component normal to the local streaming.
As time progresses, the stretching of the strips causes the length scale in the $n$ direction to become significantly smaller than the length scale in the $\sigma$ direction. As a result, we may neglect the axial advection term in the transport equation \eqref{eq:gov_eq}.
The transport equation in the Lagrangian frame, after the warped time coordinate transform, is obtained as 
\begin{equation}
\frac{\partial a}{\partial \theta} = \frac{\partial^2 a}{\partial n^2},
\end{equation}
which is subjected to the condition that at $\theta = 0$ we have an initial Gaussian distribution given by $a(n) = exp(-n^2)$. The solution for this diffusion problem is obtained as \cite{Crank}
\begin{equation}
a(n,\theta) = \frac{1}{\sqrt{1+4\theta}}\exp\left(- \frac{n^2/s^2}{1+4\theta}\right).
\label{eq:appendix_3}
\end{equation}
In this case, the maximum concentration at $n = 0$ is obtained as ${1}/{\sqrt{1+4\theta}}$ while the width of the distribution is $\sqrt{\frac{\int_{-\infty}^{\infty}n^2a(n,\theta)dn}{\int_{-\infty}^{\infty}a(n,\theta)dn}} = s \sqrt{\frac{{1+4\theta}}{2}}$.
\section*{References}
\bibliographystyle{elsarticle-num}
\bibliography{rxnstrip}
\end{document}